\documentclass[preprint,aps,12pt,nofootinbib, onecolumn,superscriptaddress,preprintnumbers,balancelastpage,longbibliography]{revtex4-1}

\usepackage{dcolumn}
\usepackage{xcolor}
\usepackage{youngtab}
\usepackage{ytableau}

\usepackage{booktabs}    
\usepackage{placeins}
\usepackage{soul}

\definecolor{redd}{rgb}{0.8, 0.1,0.2}
\definecolor{navy}{rgb}{0.05, 0.23,0.75}
\usepackage[colorlinks=true,citecolor=red,linkcolor=blue]{hyperref}
\usepackage{chngcntr}
\hypersetup{
     colorlinks   = true,
     citecolor    = navy,
	linkcolor = redd,
	urlcolor=navy,
	anchorcolor=blue
}
\usepackage{bbm}
\usepackage{amsfonts}
\usepackage{amsmath,amssymb}
\usepackage{mathrsfs}
\usepackage{epsfig}
\usepackage{graphicx}
\usepackage{wrapfig}                 
\usepackage{url}
\usepackage{hyperref}
\usepackage{float}
\usepackage{color}
\usepackage{multirow}
\usepackage{lipsum}
\usepackage{enumitem}
\usepackage{ctable}
\newcolumntype{L}{>{\centering\arraybackslash}m{1.5cm}}

\usepackage{enumitem}
\usepackage{chngcntr}

\newcommand{\nn}{\nonumber}

\newcommand{\be}{\begin{equation}}
\newcommand{\ee}{\end{equation}}
\newcommand{\bea}{\begin{eqnarray}}
\newcommand{\eea}{\end{eqnarray}}
\newcommand{\bc}{\begin{center}}
\newcommand{\ec}{\end{center}}

\begin{document}
		
\title{Staggered Fermions with Chiral Anomaly Cancellation}

\author{Ling-Xiao Xu}
\email{phy.lingxiao.xu@gmail.com}
\affiliation{Abdus Salam International Centre for Theoretical Physics, Strada Costiera 11, 34151, Trieste, Italy}

\begin{abstract}
We investigate the implications of the quantized vectorial and axial charges in the lattice Hamiltonian of multi-flavor staggered fermions in $(1+1)$ dimensions. These lattice charges coincide with those of the $U(1)_V$ and $U(1)_A$ global symmetries of Dirac fermions in the continuum limit, whose perturbative chiral anomaly matches the non-Abelian Onsager algebra on the lattice. In this note, we focus on the lattice models that flow to continuum quantum field theories of Dirac fermions that are free from the perturbative chiral anomaly between $U(1)_V$ and $U(1)_A$. 
In a lattice model that flows to two Dirac fermions, we identify quadratic Hamiltonian deformations that can gap the system while fully preserving both the vectorial and axial charges on the lattice. These deformations flow to the usual symmetry-preserving Dirac mass terms in the continuum. Additionally, we propose a lattice model that flows to the chiral fermion $3-4-5-0$ model in the continuum by using these lattice charges, and we discuss the multi-fermion interactions that can generate a mass gap in the paradigm of symmetric mass generation.
\end{abstract}

\maketitle

\tableofcontents
	
\section{Introduction}
\label{sec:intro}

One excellent question in lattice models and Quantum Field Theories (QFT) in the continuum is to understand their symmetries and anomalies, and the relations between them along the renormalization group flow. The lattice model can often be viewed as a regularization in the ultraviolet (UV) which concretely defines the theory at a fundamental level. In contrast, the QFT in the continuum only emerges as the low energy limit in the infrared (IR). Given some symmetries in the QFT in the IR, one may ask how these symmetries arise in the UV lattice model.
Furthermore, if there are 't Hooft anomalies identified in the UV lattice model, they are also expected to be matched in the QFT at IR~\cite{tHooft:1979rat}~\footnote{From a modern perspective, some classical results in condensed matter physics such as the Lieb-Schultz-Mattis theorem~\cite{Lieb:1961fr} and Luttinger theorem~\cite{Luttinger:1960ua, Luttinger:1960zz} can also be viewed as 't Hooft anomalies. See e.g.~\cite{Cheng:2022sgb} for a recent paper that presents these results in a unified fashion, and many references therein.}, even though the symmetries in the UV and IR do not have to be identical.

It is in fact very common to have different symmetries in the UV and IR. 
One familiar example is the accidental or emergent symmetries, which appear only below some energy scale. 
In contrast, there exists another class of IR symmetry known as emanant symmetries~\cite{Cheng:2022sgb, Seiberg:2023cdc, Seiberg:2024gek}. As already suggested by their name, they emanate from UV symmetries (hence not being accidental or emergent), i.e. they can be thought of as the low energy limit of some other UV symmetries satisfying a different set of algebra. 
When the emanant symmetry is exact in the low energy theory, there can be (either relevant or irrelevant) deformations that fully respect the IR emanant symmetry while violating the UV symmetry it emanates from. Furthermore, 't Hooft anomalies are expected to match for emanant symmetries but not for accidental symmetries.

Following these general considerations, a recent work~\cite{Chatterjee:2024gje} studied the Hamiltonian lattice model of staggered fermions~\cite{Kogut:1974ag, Banks:1975gq, Susskind:1976jm} which flows to one massless Dirac fermion in the continuum in $(1+1)$ dimensions. In particular, they identified the quantized lattice charges $Q^V$ and $Q^A$ emanating those of the $U(1)_V$ and $U(1)_A$ symmetries of a Dirac fermion in the continuum limit, and they also found the perturbative chiral anomaly between $U(1)_V$ and $U(1)_A$ symmetries in the continuum matches the Onsager algebra~\cite{Onsager:1943jn} on the lattice:
\be
[Q^V, Q^A] = i G_{1} \ ,
\ee
where the concrete expressions of $Q^V$, $Q^A$, and $G_1$ will be given when appropriate. Hence their work successfully realized the perturbative chiral anomaly in a lattice model which has finite-dimensional local Hilbert space. (This needs to be contrasted with other constructions on lattice anomalies~\cite{ Sulejmanpasic:2019ytl, Gorantla:2021svj, Fazza:2022fss, Berkowitz:2023pnz, 
Cheng:2022sgb,
Seifnashri:2023dpa, Catterall:2018lkj, Catterall:2022jky}.) Finally, they observe that $Q^V$ and $Q^A$ together enforce symmetry-preserving gaplessness for the lattice model, this agrees with the fact in the continuum that there is no $U(1)_V\times U(1)_A$-preserving deformation that can gap a single Dirac fermion.
(See also~\cite{Pace:2024oys} for a follow-up work on the bosonized theory.)

In this note, we take one step further and consider the lattice models that in the continuum limit flow to multiple Dirac fermions that are free from the perturbative chiral anomaly between the $U(1)_V$ and $U(1)_A$ symmetries.
After all, canceling the perturbative chiral anomaly in the continuum is straightforward by adding spectator fermions~\cite{tHooft:1979rat}. Nevertheless, it is still intriguing to understand the origin of anomaly cancellation on the lattice. 
Moreover, chiral anomaly cancellation is also well-motivated for the paradigm of symmetric mass generation (SMG); see~\cite{Wang:2022ucy} for a comprehensive review and the references therein. When chiral anomalies are canceled in the continuum (hence no obstructions to a trivially gapped phase from chiral symmetries), it is natural to ask whether there exist symmetry-preserving deformations that can drive the theory from the gapless phase to the gapped phase. Once these symmetry-preserving deformations are realized on the lattice, we are interested in understanding whether they are fully compatible with the lattice charges. 
By definition, the Hamiltonian deformations that are consistent with the lattice charges $Q^{V, A}$ must also be invariant under the $U(1)_V\times U(1)_A$ symmetry in the continuum, but not necessarily the other way around.

More specifically, for seeking scenarios with chiral anomaly cancellation, we consider multiple flavors of staggered fermions that flow to massless multi-flavor Dirac fermions at low energy. 
In the continuum, there is an Abelian $U(1)_V \times U(1)_A$ subgroup where individual Dirac fermion $\Psi_I$ carries charges $(p_I, q_I)$ with $p_I, q_I\in \mathbb{Z}$. Hence the chiral anomaly between $U(1)_V$ and $U(1)_A$ is given by $\sum_I p_I q_I$ where the sum runs over all flavors of Dirac fermions. This matches the Onsager algebra on the lattice as
\begin{align}
[Q^V,Q^A]&\equiv \left[\sum_I p_I Q^{V_I}, \sum_{I^\prime} q_{I^\prime}  Q^{A_{I^\prime}}\right]=i\sum_{I I^\prime} p_I q_{I^\prime} \delta_{I,I^\prime} G_{I,1}\nn\\
&=i\sum_{I} p_I q_I G_{I,1}\; ,
\end{align}
where $Q^{V_I}$ and $Q^{A_I}$ are the lattice charges acting only on the lattice fermion that flows to the Dirac fermion $\psi_I$.
We will focus on the scenario where the perturbative chiral anomaly vanishes, i.e. $\sum_I p_I q_I=0$, and ask whether there are deformations that gap the entire system while fully preserving both the lattice charges $Q^V$ and $Q^A$ (or only in the continuum limit with the symmetry charges that emanate from $Q^V$ and $Q^A$). We consider in this note only $I=2$ for simplicity.

The rest of the paper is organized as follows. In Section~\ref{sec:review_oneDirac} we set the stage by briefly reviewing the known results on the quantized charges of staggered fermions. In Section~\ref{sec:model_2} we consider a lattice model that flows to two Dirac fermions in the continuum limit, and then we discuss the Hamiltonian deformations that can gap the entire system while fully preserving the lattice charges. As we will see, they correspond to the symmetry-preserving Dirac mass terms. In Section~\ref{sec:model_3450} we propose a chiral fermion $3-4-5-0$ lattice model generalizing the previous results, and we discuss the multi-fermion interactions that can drive the system to a symmetry-preserving gapped phase. We find these interaction terms are only compatible with the symmetry charges in the continuum limit, where they are consistent with the paradigm of symmetric mass generation. Finally we conclude in Section~\ref{sec:conclusion} and summarize some technicalities in Appendices~\ref{app1} and~\ref{app2}.

\section{A brief review on quantized charges of staggered fermions}
\label{sec:review_oneDirac}
 
This section is meant to set the stage and mainly contains the known results on staggered fermions necessary for our analysis in later sections. 
Readers familiar with the topic may feel free to skip this section.

\subsection{Hamiltonian of staggered fermions}

In $(1+1)$ dimensional spacetime in the continuum, the action of a single Dirac fermion and the corresponding Hamiltonian are given by 
\begin{align}
\mathcal{S}&= i \int \text{d}t\text{d}x \left[ \psi^\dagger_L (\partial_t-\partial_x) \psi_L + \psi^\dagger_R (\partial_t+\partial_x) \psi_R\right]\;, \\
\mathcal{H}&=i\int \text{d}x \left[ \psi^\dagger_L \partial_x \psi_L -  \psi^\dagger_R \partial_x \psi_R\right]\;.
\label{eq:freeHam_one_Dirac}
\end{align}
Here $\psi_{L,R}$ are the left-moving and right-moving complex Weyl fermions, respectively. Notice that in the basis of $\psi_{L,R}$ the Dirac matrices are $\Gamma_0=\sigma^x, \Gamma_1=-i\sigma^y$, and $\Gamma_5\equiv \Gamma_0\Gamma_1=\sigma^z$.
It is well known that naive discretization of Eq.~\eqref{eq:freeHam_one_Dirac} leads to the fermion doubling problem; see Appendix~\ref{app1} for a review. 
 
Staggered fermions~\cite{Kogut:1974ag, Banks:1975gq, Susskind:1976jm} avoid the fermion doubling problem by having a one-component complex lattice fermion $c_j$ on each lattice site (see e.g.~\cite{Susskind:1976jm}). 
Throughout the paper, we assume that the total number of lattice sites $N$ is an even integer.   
The lattice Hamiltonian of staggered fermions reads
\be
H=\frac{i}{2} \sum^{N}_{j=1} (c^\dagger_{j} c_{j+1} + c_{j} c^\dagger_{j+1} )\ ,
\label{eq:Ham_staggered}
\ee
where the one-component complex lattice fermions $c_j$ satisfy the usual Clifford algebra and they flow to the left-mover $\psi_{L}$ and the right-mover $\psi_R$ in the continuum. This is most easily seen in the momentum space where $c_j$ is related to its 
counterpart $\gamma_k$ as $c_j=\frac{1}{\sqrt{N}}\sum_{k} e^{\frac{2\pi i}{N} k j } \gamma_{k}$, and accordingly the Hamiltonian in the momentum space is
\be
H=-\sum_{k} \sin\left(\frac{2\pi}{N} k\right) \gamma^\dagger_{k} \gamma_{k}
\label{eq:Ham_staggered_momentum}
\ee
where $\psi_L$ (and its conjugate $\psi^\dagger_L$) in the continuum is identified as the low-energy mode near $k=0$, while $\psi_R$ (and its conjugate $\psi^\dagger_R$) in the continuum is identified as the low-energy mode near $k=\pm \frac{N}{2}$ (where $k$ and $k+N$ are identified, i.e. $k\sim k+N$). More specifically, $\psi^\dagger_L(-|\epsilon|)$ corresponds to $\gamma^\dagger_{-|\epsilon|}$, $\psi_L(-|\epsilon|)$ corresponds to $\gamma_{|\epsilon|}$, $\psi^\dagger_R(|\epsilon|)$ corresponds to $\gamma^\dagger_{-\frac{N}{2}+|\epsilon|}$, and $\psi_R(|\epsilon|)$ corresponds to $\gamma_{\frac{N}{2}-|\epsilon|}$, where for low-energy excitations $|\epsilon|\ll\frac{N}{2}$.

The state with the lowest energy defines the vacuum $|\Omega\rangle$. For the staggered fermion Hamiltonian $H$, it is defined by filling all the negative energy modes while all the positive energy modes are unfilled. Therefore, it requires that for the Hamiltonian in Eq.~\eqref{eq:Ham_staggered_momentum},
$\gamma_k |\Omega\rangle= \gamma^\dagger_{-k}|\Omega\rangle = 0$ where $k\in (-\frac{N}{2},0)$.
The complex lattice fermions $c_j$ are subject to the boundary condition (B.C.) $c_{j+N}=(-1)^\nu c_j$, where periodic B.C. corresponds $\nu=0$ and anti-periodic B.C. corresponds $\nu=1$. This implies the quantization condition for the momentum to be $k=\frac{\nu}{2}+\mathbb{Z}$. For periodic B.C. we have fermion zero modes at $k=0$ and $k=\pm \frac{N}{2}$, and there are four degenerate vacua, i.e. they are $|\Omega\rangle, \gamma^\dagger_0|\Omega\rangle, \gamma^\dagger_{\frac{N}{2}}|\Omega\rangle$ and $\gamma^\dagger_0\gamma^\dagger_{\frac{N}{2}}|\Omega\rangle$. On the other hand, for anti-period B.C. there is only one unique vacuum $|\Omega\rangle$.

\subsection{Quantized charges and a chiral anomaly on lattice}

In the continuum, the theory of one massless, free Dirac fermion enjoys the $U(1)_L\times U(1)_R$ chiral symmetry, which can be put in a linear combination using vectorial and axial symmetries $U(1)_V\times U(1)_A$~\footnote{Here we neglect the global structure of global symmetry group $U(1)_V\times U(1)_A$, but we note that there is a $\mathbb{Z}_2$ subgroup generated by the group element $(e^{i\pi \mathcal{Q}^V}, e^{i\pi \mathcal{Q}^A})$ acting trivially on both $\psi_L$ and $\psi_R$.}, whose generators can be denoted as $\mathcal{Q}^V$ and $\mathcal{Q}^A$.
They act on chiral fermions $\psi_L$ and $\psi_R$ as follows,
\be
[\mathcal{Q}^V, \psi^\dagger_L]=\psi^\dagger_L\;, \quad
[\mathcal{Q}^V, \psi^\dagger_R]=\psi^\dagger_R\;,\quad
[\mathcal{Q}^A, \psi^\dagger_L]=\psi^\dagger_L\;, \quad 
[\mathcal{Q}^A, \psi^\dagger_R]=-\psi^\dagger_R\;,
\label{eq:conti_action}
\ee
where $\mathcal{Q}^V=\mathcal{Q}^L+\mathcal{Q}^R$ and $\mathcal{Q}^A=\mathcal{Q}^L-\mathcal{Q}^R$. 
It is important to notice that the vectorial and axial charges in the continuum are related by conjugating the charge conjugation operator $\mathcal{C}^R$, 
\be
\mathcal{Q}^A=\mathcal{C}^R \mathcal{Q}^V (\mathcal{C}^R)^{-1}\; ,
\label{eq:conti_axial_charge}
\ee
where $\mathcal{C}^R$ acts on $\psi_R$ as $\mathcal{C}^R \psi_R (\mathcal{C}^R)^{-1}= \psi_R^\dagger$ and hence $\mathcal{C}^R \mathcal{Q}_R (\mathcal{C}^R)^{-1}=-\mathcal{Q}_R$.~\footnote{Because of this algebra, the chiral $U(1)_R$ group is extended to $O(2)_R=U(1)_R \rtimes (\mathbb{Z}_2)_R$. Similarly, one also has the charge conjugation operator $\mathcal{C}^L$ acting on $\psi_L$ as $\mathcal{C}^L \psi_L(\mathcal{C}^L)^{-1}=\psi_L^\dagger$ and $\mathcal{C}^L\mathcal{Q}_L(\mathcal{C}^L)^{-1}=-\mathcal{Q}_L$.} 
There is a perturbative mixed chiral anomaly between $U(1)_V$ and $U(1)_A$ symmetries, which can be captured by Feynman diagrams and is understood to be the obstruction of simultaneously gauging $U(1)_V$ and $U(1)_A$. Because of this anomaly, the theory of a single massless Dirac fermion cannot be trivially gapped while preserving both $\mathcal{Q}^V$ and $\mathcal{Q}^A$. 

Next we discuss the lattice precursors of $\mathcal{Q}^V$ and $\mathcal{Q}^A$ following~\cite{Chatterjee:2024gje}. To construct the lattice charges, the authors of~\cite{Chatterjee:2024gje} showed that it is useful to work with the lattice Majorana fermions, where they are related to the complex lattice fermions $c_j$ as $c_j=\frac{1}{2} \left( a_j+i b_j \right)$.
One can deduce that $a_j, b_j$ satisfy the algebra $\{a_j,a_{j^\prime}\}=\{b_j,b_{j^\prime}\}=2 \delta_{j,j^\prime}$ and $\{a_j,b_{j^\prime}\}=0$. In the continuum limit, they respectively flow to a pair of Majorana-Weyl fermions. We refer the readers to~\cite{Chatterjee:2024gje} for more detailed discussions. For our purposes, we note that in terms of the Majorana lattice fermions the staggered fermion Hamiltonian reads
\be
H=\frac{i}{4} \sum_{j=1}^N (a_j a_{j+1}+b_j b_{j+1})\;.
\ee
One can observe that the fermion number 
\be
Q^V=\sum_{j=1}^{N} \left(c^\dagger_j c_j-\frac{1}{2}\right)=\sum_{j=1}^{N}\frac{i}{2} a_j b_j
\label{eq:vec_lat_charge}
\ee
is preserved by the staggered fermion Hamiltonian.
Furthermore, there is another conserved lattice charge obtained by conjugating $Q^V$ with the lattice translation operator $T_b$ for Majorana fermions $b_j$, 
\bea
Q^A&=&(T_b)^\delta \ Q^V \ (T_b)^{-\delta}\label{eq:axi_lat_charge_1}\\
&=&\frac{1}{2}\sum_{j=1}^{N}(c_j+c^\dagger_j)(c_{j+\delta}-c^\dagger_{j+\delta})=\frac{i}{2}\sum_{j=1}^N a_j b_{j+\delta}\;,
\label{eq:axi_lat_charge}
\eea
where $T_b$ acts on $b_j$ as $T_b b_j (T_b)^{-1}=b_{j+1}$ while leaves $a_j$ invariant.~\footnote{Similarly, one can have the lattice translation operator for $a_j$.} Here the lattice span $\delta$ is required to be an odd integer much smaller than the total number of lattice sites (i.e. $\delta\ll N$). We will set $\delta=1$ for simplicity. We will see that $Q^V$ and $Q^A$ respectively flow to $\mathcal{Q}^V$ and $\mathcal{Q}^A$ in the continuum. 

In the above construction of $Q^A$, the lattice translation operator $T_b$ has played a crucial role. One might be curious to find an explicit expression for it. Following~\cite{Seiberg:2023cdc}, for an even number of lattice sites, we have the lattice translation operator for a single flavor Majorana fermion $b_j$ as follows,  
\be
\tilde{T}_{b}=\frac{1}{2^{\frac{N-1}{2}}} b_1 (1+b_1 b_2) (1+b_2 b_3) \cdots (1+b_{N-1} b_N)\ .
\ee
It acts on $b_j$ properly, i.e. $\tilde{T}_b b_j (\tilde{T}_b)^{-1}=b_{j+1}$, but it does not leave $a_j$ invariant, instead we find $\tilde{T}_b a_j (\tilde{T}_b)^{-1}=-a_j$. This minus sign can be fixed by using the lattice parity operator for the Majorana fermion $a_j$~\cite{Seiberg:2023cdc} 
\be
\tilde{G}_a=a_1a_2\cdots a_N\;, 
\ee
it acts on $a_j$ as $\tilde{G}_a a_j (\tilde{G}_a)^{-1}=-a_j$ while it leaves $b_j$ invariant. 
Therefore, we find that the operator acting properly on both $a_j$ and $b_j$ is $T_b\equiv \tilde{T}_{b} \tilde{G}_a $.~\footnote{This construction can easily be generalized when there are more flavors of lattice Majorana fermions.}
Here the insight in the construction of $Q^A$ is that the lattice translation operator $T_b$ emanates the charge conjugation operator $\mathcal{C}^R$ in the continuum~\cite{Seiberg:2023cdc}. Hence the lattice charge in Eq.~\eqref{eq:axi_lat_charge_1} has the same form as its continuum counterpart in Eq.~\eqref{eq:conti_axial_charge}.

One can go to the momentum space and check how $Q^V$ and $Q^A$ act, the result is
\begin{align}
[Q^V, \gamma^\dagger_k] &=\gamma^\dagger_k\;,\\
[Q^A, \gamma^{\dagger}_k] &= \cos\left(\frac{2\pi }{N} k\right) \gamma^{\dagger}_k + i \sin\left(\frac{2\pi }{N} k\right) \gamma_{-k}\;.
\label{eq:lat_action}
\end{align}
We see that $Q^V$ acts in a vectorial fashion while $Q^A$ acts in an axial fashion on the low-energy modes near $k=0$ and $k=\pm \frac{N}{2}$. In particular, by the fermion correspondence $\psi^\dagger_L(-|\epsilon|)\sim \gamma^\dagger_{-|\epsilon|}$ and $\psi^\dagger_R(|\epsilon|)\sim \gamma^\dagger_{-\frac{N}{2}+|\epsilon|}$ and taking the continuum limit (i.e. $|\epsilon|\ll N$ and $N\to\infty$), it is easy to see that the actions of $Q^V$ and $Q^A$ coincide with the actions of $\mathcal{Q}^V$ and $\mathcal{Q}^A$ in the continuum, c.f. Eq.~\eqref{eq:conti_action}. 
There are some interesting features noted in~\cite{Chatterjee:2024gje}: 
\begin{enumerate}
    \item Although $Q^V$ and $Q^A$ are exact lattice symmetries for Eq.~\eqref{eq:Ham_staggered}, they do not commute (rather they satisfy the Onsager algebra), hence one cannot define fermion chirality based on $(Q^V\pm Q^A)/2$, i.e. the existence of $Q^V$ and $Q^A$ has no contradiction to the Nielsen-Ninomiya theorem~\cite{Nielsen:1980rz, Nielsen:1981hk, Nielsen:1981xu}.
Moreover, the Onsager algebra is matched by the chiral anomaly in the continuum.

    \item Another feature is that $Q^V$ and $Q^A$ together enforce the gaplessness of the lattice model. This is consistent with the continuum QFT, where for a single Dirac fermion there is a mixed chiral anomaly obstructing a mass gap. 
\end{enumerate}

\section{A vectorlike model of two Dirac fermions}
\label{sec:model_2}

In this section, we take a step further and analyze the quantized charges in a model of two Dirac fermions both in the continuum and on the lattice. In particular, we will focus on an anomaly-free Abelian group and identify the symmetry-preserving deformations that can gap the entire theory. 

\subsection{In the continuum}

In the continuum the Hamiltonian of two Dirac fermions $\psi_1$ and $\psi_2$ is given by
\be
\mathcal{H}=i\sum_{I}\int \text{d}x \left[ \psi^\dagger_{I,L} \partial_x \psi_{I,L} -  \psi^\dagger_{I,R} \partial_x \psi_{I,R}\right]\; , 
\label{eq:ham_model2_conti}
\ee
where the fermion flavor index $I=1,2$. The model has the Abelian symmetry group~\footnote{As in the previous model of a single Dirac fermion, we neglect the global structure of $G_{\text{conti.}}$. Moreover, we note that the full chiral symmetry group is $U(2)_L\times U(2)_R$, but here we only focus on the Abelian subgroup $G_{\text{conti.}}$ and its conserved charges. We leave it to a future study on the conserved non-Abelian charges.}
$G_{\text{conti.}}=U(1)_{V_1}\times U(1)_{A_1}\times U(1)_{V_2}\times U(1)_{A_2}$, where $(\psi_1)_{L,R}$ are charged under $U(1)_{V_1}\times U(1)_{A_1}$ and $(\psi_2)_{L,R}$ are charged under $U(1)_{V_2}\times U(1)_{A_2}$. For our purpose, we are particularly interested in an anomaly-free subgroup of $G_{\text{conti.}}$ and its symmetry-preserving deformations. The anomaly-free and Abelian subgroup is
\be
G^\prime_{\text{conti.}}=U(1)_V\times U(1)_A \;,
\ee
whose charge assignments are summarized in Table~\ref{tab:model_2Dirac}. Indeed one can check the perturbative chiral anomalies of $U(1)_V\times U(1)_A$ cancel. Furthermore, since the number of left movers matches that of right movers, there is no mixed anomaly with gravity either.
We denote the symmetry charges in the continuum for the anomaly-free $U(1)_V$ and $U(1)_A$ as
\bea
\mathcal{Q}^V &=&  \mathcal{Q}^{V_1}+\mathcal{Q}^{V_2}\ ,\\
\mathcal{Q}^A &=& \mathcal{Q}^{A_1}- \mathcal{Q}^{A_2}\ ,
\eea
where $\mathcal{Q}^{V_I}$ and $\mathcal{Q}^{A_I}$ acts on the corresponding fermions $(\psi_I)_{L,R}$ in the same fashion as in Eq.~\eqref{eq:conti_action}. Therefore we have
\begin{align}
[\mathcal{Q}^V, \psi^\dagger_{1,L}]&=\psi^\dagger_{1,L}\;, \quad 
[\mathcal{Q}^V, \psi^\dagger_{1,R}]=\psi^\dagger_{1,R}\;, \quad
[\mathcal{Q}^V, \psi^\dagger_{2,L}]=\psi^\dagger_{2,L}\;, \quad 
[\mathcal{Q}^V, \psi^\dagger_{2,R}]=\psi^\dagger_{2,R}\; ;\nn\\
[\mathcal{Q}^A, \psi^\dagger_{1,L}]&=\psi^\dagger_{1,L}\;, \quad
[\mathcal{Q}^A, \psi^\dagger_{1,R}]=-\psi^\dagger_{1,R}\;, \quad
[\mathcal{Q}^A, \psi^\dagger_{2,L}]=-\psi^\dagger_{2,L}\;, \quad
[\mathcal{Q}^A, \psi^\dagger_{2,R}]=\psi^\dagger_{2,R}\;.
\label{eq:conti_action_model2}
\end{align}
Since all the charges $\mathcal{Q}^{V_I}$ and $\mathcal{Q}^{A_I}$ commute, both fermion flavor and chirality are well-defined notions in the continuum. 

\begin{table}[t]
      \centering
      \begin{tabular}{|c||c|c||c|}
      \hline
           & $U(1)_V$ & $U(1)_A$ & \\
           \hline
         $\psi^\dagger_{1L}$  & $1$ & $1$ & $\zeta^\dagger_{1L}$\\
         \hline
         $\psi^\dagger_{1R}$  & $1$  & $-1$ & $\zeta^\dagger_{2R}$ \\
         \hline
         $\psi^\dagger_{2L}$  & $1$ & $-1$ & $\zeta^\dagger_{2L}$\\
         \hline
         $\psi^\dagger_{2R}$  & $1$ & $1$ & $\zeta^\dagger_{1R}$\\
         \hline
      \end{tabular}
      \caption{Charge assignments of an anomaly-free $U(1)_V\times U(1)_A$ symmetry group in a model of two Dirac fermions in the continuum. In particular, since $U(1)_V\times U(1)_A$ is free of chiral anomalies, there is no obstruction to a trivially gapped phase while fully preserving the $U(1)_V\times U(1)_A$ symmetry. The model seems chiral in the $\Psi_{I}=(\psi_{I,L},\psi_{I,R})^T$ basis (where $I=1,2$), but a simple relabeling of fields in terms of $\zeta_{I}=(\zeta_{I,L},\zeta_{I,R})^T$ renders the vectorlike nature of the model manifest. Namely, for a left-mover, there must be a right-mover with the same quantum numbers, such that symmetry-preserving mass terms exist.
      }
      \label{tab:model_2Dirac}
\end{table}

In the continuum, it is intuitive to notice that the Dirac mass terms (i.e. fermion bilinear operators) in the form of $\overline{\Psi}_1\Psi_2$ or $\overline{\Psi}_1\Gamma_5\Psi_2$ preserve $G^\prime_{\text{conti.}}=U(1)_V\times U(1)_A$ and these two terms can gap the Dirac fermions $\Psi_1$ and $\Psi_2$, here we define $\Psi_{I}=(\psi_{I,L},\psi_{I,R})^T$. The corresponding Hamiltonian deformation reads
\begin{align}
\delta \mathcal{H}&= \int \text{d}x \left(m_1
\psi^\dagger_{1,L}\psi_{2,R}+m_2 \psi^\dagger_{1,R}\psi_{2,L}+\text{h.c.}\right) \\
&=\int \text{d}x \left(g_1 \overline{\Psi}_1\Psi_2+g_2\overline{\Psi}_1\Gamma_5\Psi_2+\text{h.c.}\right)\;,
\label{eq:Ham_def_conti}
\end{align}
where $m_{1,2}$ are two independent parameters and $g_{1,2}=(\pm m_1+m_2)/2$ accordingly.~\footnote{We note that $g_{1,2}$ in general are complex parameters. However, only the relative phase between $g_1$ and $g_2$ are physical, while the overall phase can be removed by field redefinitions of either $\Psi_1$ or $\Psi_2$.} Indeed one needs two independent deformations to explicitly break $G_{\text{conti}.}$ to $G^\prime_{\text{conti}.}$, and the anomaly-free subgroup $G^\prime_{\text{conti}.}$ cannot obstruct the mass gap generated by $\delta \mathcal{H}$, which only includes the $G^\prime_{\text{conti}.}$-preserving terms.

\subsection{On the lattice}
  
Next we analyze the corresponding conserved charges in a lattice model of two flavors of staggered fermions which in the continuum limit flows to the model of two Dirac fermions. 
The Hamiltonian of two flavors of  staggered fermions is given by
\begin{align}
H&=\frac{i}{2}\sum_{I} \sum^{N}_{j=1} (c^\dagger_{I, j} c_{I, j+1} + c_{I, j} c^\dagger_{I, j+1} )\nn\\
&=\frac{i}{4} \sum_{I} \sum_{j=1}^N (a_{I,j} a_{I,j+1}+b_{I,j} b_{I,j+1})\;,
\label{eq:ham_model2}
\end{align}
where $I=1,2$ is the flavor index. Here the complex lattice fermions satisfy the Clifford algebra $\{c_{I,j},c^\dagger_{I^\prime,j^\prime}\}=\delta_{j,j^\prime}\delta_{I,I^\prime}$ and $\{c_{I,j},c_{I^\prime,j^\prime}\}=\{c^\dagger_{I,j},c^\dagger_{I^\prime,j^\prime}\}=0$. Likewise, $a_{I,j}$ and $b_{I,j}$ are the lattice Majorana fermions that are obtained as $a_{I,j}=c_{I,j}+c^\dagger_{I,j}$ and $b_{I,j}=-i(c_{I,j}-c^\dagger_{I,j})$. One can deduce the algebra satisfied by $a_{I,j}$ and $b_{I,j}$ as the following: $\{a_{I,j},a_{I^\prime,j^\prime}\}=\{b_{I,j},b_{I^\prime,j^\prime}\}=2\delta_{I,I^\prime}\delta_{j,j^\prime}$ and $\{a_{I,j},b_{I^\prime,j^\prime}\}=0$.

In analog to the model of a single Dirac fermion, we have the charges for conserved fermion numbers respectively for $c_{1,j}$ and $c_{2,j}$ which can be denoted as $Q^{V_1}$ and $Q^{V_2}$, and they respectively flow to $\mathcal{Q}^{V_1}$ and $\mathcal{Q}^{V_2}$ in the continuum limit. It is intuitive to define the lattice charge that emanates $\mathcal{Q}^V$ in the continuum:
\begin{align}
Q^V=Q^{V_1}+Q^{V_2}&=\sum_{j=1}^{N} \left(c^\dagger_{1,j} c_{1,j}-\frac{1}{2}\right)+\sum_{j=1}^{N} \left(c^\dagger_{2,j} c_{2,j}-\frac{1}{2}\right)\\
&=\frac{i}{2} \sum_{j=1}^N (a_{1,j} b_{1,j}+a_{2,j} b_{2,j})\; ,
\label{eq:vec_lat_charge_model2}
\end{align}
Furthermore, one can define the lattice axial charge that emanates $\mathcal{Q}^A$ in the continuum: 
\begin{align}
Q^A&=Q^{A_1}-Q^{A_2}=(T_{b_1})^{\delta_1} \ Q^{V_1} \ (T_{b_1})^{-\delta_1}-(T_{b_2})^{\delta_2} \ Q^{V_2} \ (T_{b_2})^{-\delta_2}\\
&=\frac{1}{2}\sum_{j=1}^{N}(c_{1,j}+c^\dagger_{1,j})(c_{1,j+\delta_1}-c^\dagger_{1,j+\delta_1})-\frac{1}{2}\sum_{j=1}^{N}(c_{2,j}+c^\dagger_{2,j})(c_{2,j+\delta_2}-c^\dagger_{2,j+\delta_2})\\
&=\frac{i}{2}\sum_{j=1}^N (a_{1,j} b_{1,j+\delta_1}-a_{2,j} b_{2,j+\delta_2})\;,
\label{eq:axi_lat_charge_model2}
\end{align}
where $\delta_{1,2}$ are odd integers much smaller than total number of lattice sites, and the lattice translation operators $T_{b_I}$ acts on the lattice Majorana fermion $b_I$ as $T_{b_I} b_{I,j} T^{-1}_{b_I}=b_{I,j+1}$ while leaving all other Majorana fermions invariant. As we mentioned earlier, it is important to use the lattice parity operator of Majorana fermions in the construction of various $T_{b_I}$. 

Clearly, since $[Q^{V_1}, Q^{V_2}]=[Q^{A_1}, Q^{A_2}]=[Q^{V_1}, Q^{A_2}]=[Q^{V_2}, Q^{A_1}]=0$, fermion flavor remains as a well-defined notion on the lattice. However, since $[Q^{V_I}, Q^{A_I}]\neq 0$ for both $I=1,2$, fermion chirality is not a well-defined notion on the lattice. Indeed this is consistent with the Nielsen-Ninomiya theorem. 
Following~\cite{Chatterjee:2024gje}, it is intriguing to evaluate the following commutator 
\be
[Q^V,Q^A]=i(G_{1,\delta_1}-G_{2,\delta_2})\ ,
\label{eq:Latticeanomaly_2Dirac}
\ee
where $G_{I,\delta_I}=\frac{i}{2}\sum_j(a_{I,j}a_{I,j+\delta_I}-b_{I,j}b_{I,j+\delta_I})$. The relative minus sign between $G_{1,\delta_1}$ and $G_{2,\delta_2}$ implies chiral anomaly cancellation between two fermion flavors. This matches the results in the continuum limit, indeed there is no mixed anomaly between $U(1)_V$ and $U(1)_A$ for charge assignments in Table~\ref{tab:model_2Dirac}.
Of course all the matrix elements of $G_{I,\delta_I}$ between two low-energy states vanish in the continuum limit~\cite{Chatterjee:2024gje}.
For simplicity, we will take $\delta_1=\delta_2=1$ in the following analysis. 

It is straightforward to check how $Q^V$ and $Q^A$ act on the lattice complex fermions in the momentum space, i.e.
\begin{align}
[Q^V, \gamma^\dagger_{1,k}] &=\gamma^\dagger_{1,k}\;,\label{eq:lat_action_model2_1}\\
[Q^V, \gamma^\dagger_{2,k}] &=\gamma^\dagger_{2,k}\;,\label{eq:lat_action_model2_2}\\
[Q^A, \gamma^{\dagger}_{1,k}] &= \cos\left(\frac{2\pi }{N} k\right) \gamma^{\dagger}_{1,k} + i \sin\left(\frac{2\pi }{N} k\right) \gamma_{1,-k}\;,\label{eq:lat_action_model2_3}\\
[Q^A, \gamma^{\dagger}_{2,k}] &= -\cos\left(\frac{2\pi }{N} k\right) \gamma^{\dagger}_{2,k} - i \sin\left(\frac{2\pi }{N} k\right) \gamma_{2,-k}\;.
\label{eq:lat_action_model2_4}
\end{align}
Likewise, one can work out the various commutation relations $[Q^{V,A},\gamma_{I,k}]$ by conjugating the above equations. 
With the identification between the fermions in the continuum and low energy modes near $k=0$ and $k=\pm \frac{N}{2}$ on the lattice (i.e.
$\psi^\dagger_{I,L}(-|\epsilon|)\sim \gamma^\dagger_{I,-|\epsilon|}$ and $\psi^\dagger_{I,R}(|\epsilon|)\sim \gamma^\dagger_{I,-\frac{N}{2}+|\epsilon|}$), we observe that the actions of $Q^V$ and $Q^A$ on the lattice coincide with the actions of $\mathcal{Q}^V$ and $\mathcal{Q}^A$ in the continuum (see Eq.~\eqref{eq:conti_action_model2}) after taking the limit $|\epsilon|\ll N$ and $N\to\infty$.

Next we discuss Hamiltonian deformations that flow to the symmetry-preserving Dirac mass terms in the continuum. Given the strong constraint on the gaplessness imposed by $Q^V$ and $Q^A$~\cite{Chatterjee:2024gje} in the model of only one flavor of staggered fermion, it is interesting to understand whether there exists any lattice Hamiltonian deformations which can gap the system. At the same time, they preserve both lattice charges $Q^V$ and $Q^A$.

Following~\cite{Susskind:1976jm, Banks:1975gq}, we first consider the lattice operator $(-1)^j c^\dagger_{1,j} c_{2,j}$ that is supposed to flow to $\overline{\Psi}_1\Psi_2$ in the continuum. Let us consider the Hamiltonian deformation 
\be
\Delta H_1=\sum_{j=1}^{N} (-1)^j c^\dagger_{1,j} c_{2,j}=\sum_{k} \gamma^\dagger_{1,k+\frac{N}{2}}\gamma_{2,k}\;.
\ee
Indeed $\Delta H_1$ contains the terms for the low energy modes $\gamma^\dagger_{1,\frac{N}{2}}\gamma_{2,0}+\gamma^\dagger_{1,0}\gamma_{2,\frac{N}{2}}$, which corresponds to the Dirac mass term $\psi^\dagger_{1,R}\psi_{2,L}+\psi^\dagger_{1,L}\psi_{2,R}$ in the continuum. Notice that we have used the identification $k\sim k+N$.
Since $\Delta H_1$ conserves the total fermion number, it commutes with $Q^V$, i.e. $[Q^V, \Delta H_1]=0$. On the other hand, it is less trivial to check whether $\Delta H_1$ commutes with $Q^A$. By straightforward calculation we find
\be
[Q^A,\Delta H_1]=i \sum_k \sin\left(\frac{2\pi}{N} k\right)\left(\gamma_{2,-k-\frac{N}{2}}\gamma_{1,k} + \gamma^\dagger_{2,k+\frac{N}{2}}\gamma^\dagger_{1,-k}\right)\;.
\ee
This commutator does not vanish exactly but vanishes when $k= 0$ and $k= \pm\frac{N}{2}$. This means that it vanishes for the low-energy modes in the continuum limit, i.e., the limit where the momenta of the excitations deviate from $0$ or $\pm\frac{N}{2}$ by a finite amount much smaller than $N$ and then $N$ is sent to infinity. This result matches the fact that the Dirac mass term $\overline{\Psi}_1\Psi_2$ is a $U(1)_V\times U(1)_A$-symmetric deformation in the continuum. Clearly, a mass gap can be opened by this Dirac mass term. The existence of a symmetry-preserving deformation (in the continuum limit) that can fully gap the system implies chiral anomaly cancellation. 

Furthermore, there is a stronger result: one can identify a deformation that fully preserves the lattice charges $Q^V$ and $Q^A$.
We note that $\Delta H_1$ itself is not hermitian, but we can construct two hermitian terms as $\Delta H_1+\Delta H^\dagger_1$ and $i(\Delta H_1-\Delta H^\dagger_1)$. We find  
\begin{align}
[Q^A, \Delta H_1+\Delta H^\dagger_1]&= 2\  [Q^A, \Delta H_1]\neq 0\;,\\
[Q^A, i(\Delta H_1-\Delta H^\dagger_1)]&=0\ .
\end{align}
Therefore, we conclude that the hermitian operator 
\be
i\left(\sum_{j=1}^{N} (-1)^j c^\dagger_{1,j} c_{2,j}-\text{h.c.}\right)
\label{eq:deformation1_model2}
\ee
preserves both $Q^V$ and $Q^A$ on the lattice.~\footnote{We thank Andrea Luzio for cross-checking the result and helpful discussions on this point.} In the continuum it flows to the Dirac mass term with $g_1$ being a purely imaginary coupling while $g_2=0$ in Eq.~\eqref{eq:Ham_def_conti}. Again, the presence of a Hamiltonian deformation which preserves both lattice charges $Q^V$ and $Q^A$ while generating a mass gap for the system implies that there is no mixed chiral anomaly between $\mathcal{Q}^V$ and $\mathcal{Q}^A$ in the continuum. This matches the Onsager algebra on the lattice, c.f. Eq.~\eqref{eq:Latticeanomaly_2Dirac}.

Inspired by~\cite{Susskind:1976jm}, we consider the lattice operator $(-1)^j (c^\dagger_{1,j} c_{2,j+1}-c^\dagger_{1,j+1} c_{2,j})$ that is supposed to flow to the pseudo density operator $\overline{\Psi}_1\Gamma_5\Psi_2$ in the continuum. Similar to the discussion above, we would like to understand whether it can lead to a Hamiltonian deformation which preserves both $Q^V$ and $Q^A$. Let us consider the lattice Hamiltonian deformation 
\begin{align}
\Delta H_2&=\sum_{j=1}^N \frac{(-1)^j}{2} (c^\dagger_{1,j} c_{2,j+1}-c^\dagger_{1,j+1} c_{2,j})\\
&=\sum_k \cos\left(\frac{2\pi}{N} k\right) \gamma^\dagger_{1,k+\frac{N}{2}} \gamma_{2,k}\;,
\end{align}
which contains the following interactions between the low energy modes $\psi^\dagger_{1,L}\psi_{2,R}-\psi^\dagger_{1,R}\psi_{1,L}$, where the relative minus sign is induced by the $\cos(\frac{2\pi}{N}k)$ term. Indeed this is nothing but the $\overline{\Psi}_1\Gamma_5\Psi_2$ operator in the continuum; see in Eq.~\eqref{eq:Ham_def_conti}. Since $\Delta H_2$ conserves the total fermion number, it commutes with $Q^V$. Furthermore, one can also check the commutator with $Q^A$ by straightforward calculation. We find
\be
[Q^A,\Delta H_2]=i \sum_k \sin\left(\frac{2\pi}{N} k\right) \cos\left(\frac{2\pi}{N} k\right)\left(\gamma_{2,-k-\frac{N}{2}}\gamma_{1,k} + \gamma^\dagger_{2,k+\frac{N}{2}}\gamma^\dagger_{1,-k}\right)\;,
\ee
which does not vanish exactly but indeed vanishes in the continuum limit, where $k$ deviates from $0$ or $\pm\frac{N}{2}$ by a finite amount infinitesimal compared to $N$ and then $N\to\infty$. 
Much like the previous case, we have the following commutator for the hermitian operators:
\begin{align}
[Q^A, \Delta H_2+\Delta H^\dagger_2]&= 2\  [Q^A, \Delta H_2]\neq 0\;,\\
[Q^A, i(\Delta H_2-\Delta H^\dagger_2)]&=0\ .
\end{align}
Consequently, we conclude that the hermitian operator 
\be
i\left(\sum_{j=1}^N \frac{(-1)^j}{2} (c^\dagger_{1,j} c_{2,j+1}-c^\dagger_{1,j+1} c_{2,j})-\text{h.c.}\right)
\label{eq:deformation2_model2}
\ee
preserves both lattice charges $Q^V$ and $Q^A$. The operator flows to the Dirac mass term in the continuum with the coupling $g_2$ being purely imaginary while $g_1$ vanishing; see Eq.~\eqref{eq:Ham_def_conti}. Again, the presence of this lattice-symmetry-preserving Hamiltonian deformation implies chiral anomaly cancellation between $\mathcal{Q}^V$ and $\mathcal{Q}^A$ in the continuum, and this Hamiltonian deformation can fully gap the two Dirac fermions. 

All the results in this subsection follow from the commutation relations from Eq.~\eqref{eq:lat_action_model2_1} to Eq.~\eqref{eq:lat_action_model2_4}. In the next subsection, we will adopt a different approach to prove the same results using lattice Majorana fermions. In particular, we will prove that the Hamiltonian deformations $i(\Delta H_1-\Delta H^\dagger_1)$ and $i(\Delta H_2-\Delta H^\dagger_2)$ are indeed the ones that preserve both charges $Q^V$ and $Q^A$ on the lattice, while the other two combinations $\Delta H_1+\Delta H^\dagger_1$ and $\Delta H_2+\Delta H^\dagger_2$ are not. Nevertheless, notice that all of these deformations are consistent with $U(1)_V\times U(1)_A$ symmetry in the continuum limit. 
This is consistent with our general expectation, i.e., the Hamiltonian deformations that are consistent with the lattice charges $Q^V$ and $Q^A$ must be invariant under $U(1)_V\times U(1)_A$ symmetry in the continuum, but not necessarily the other way around.

\subsection{Proof of symmetry-preserving quadratic Hamiltonian on the lattice}
\label{sec:proof_model2}
In this subsection, we would like to understand the lattice-symmetry-preserving Hamiltonian deformations from a different perspective. In particular, we will use lattice Majorana fermions and the translation operators. The analysis here is motivated by the strong claim in~\cite{Chatterjee:2024gje}, which states a result that $Q^V$ and $Q^A$ together obstruct a mass gap for the lattice model of staggered fermions which flows to a single Dirac fermion in the continuum limit. This obstruction is consistent with the Onsager algebra, which is regarded as the lattice precursor of the chiral anomaly between $\mathcal{Q}^V$ and $\mathcal{Q}^A$ in the continuum. Here we will show that lattice charges $Q^V$ and $Q^A$ defined in Eqs.~\eqref{eq:vec_lat_charge_model2} and~\eqref{eq:axi_lat_charge_model2} together cannot forbid the Hamiltonian deformations in Eqs.~\eqref{eq:deformation1_model2} and~\eqref{eq:deformation2_model2}, hence signaling chiral anomaly cancellation in the model of two Dirac fermions in the continuum.

With the lattice charges $Q^V=Q^{V_1}+Q^{V_2}$ and $Q^A=Q^{A_1}-Q^{A_2}$, one can construct the following lattice translation operator which eventually can be expressed in terms of that of various Majorana fermions, 
\be
T\equiv e^{-i \frac{\pi}{2} Q^A} e^{i \frac{\pi}{2} Q^V}=\left(e^{-i \frac{\pi}{2} Q^{A_1}} e^{i \frac{\pi}{2} Q^{V_1}}\right) \left(e^{i \frac{\pi}{2} Q^{A_2}} e^{i \frac{\pi}{2} Q^{V_2}}\right)
= \left(T_{b_1} T_{a_1}^{-1}\right) G_{2}\left(T_{b_2}T_{a_2}^{-1}\right)\;,
\label{eq:latticeTrans_model2}
\ee
where $G_2$ is the lattice fermion parity operator $G_2=\exp(i\pi Q^{A_2})$,
and it acts on $a_2$ and $b_2$ by conjugation as $G_2 a_{2,j} (G_2)^{-1}=-a_{2,j}$ and $G_2 b_{2,j} (G_2)^{-1}=-b_{2,j}$ while it leaves $a_{1,j}$ and $b_{1,j}$ invariant. 
Therefore the operator $T$ acts on various lattice Majorana fermions as
\be
T a_{1,j} T^{-1}= a_{1,j-1}\;, 
\; T b_{1,j} T^{-1}= b_{1,j+1}\;, 
\; T a_{2,j} T^{-1}= -a_{2,j-1}\;, 
\;T b_{2,j} T^{-1}= -b_{2,j+1}\;.
\ee

The invariance under $T$ actions already restricts the Hamiltonian deformations significantly. Similar to the arguments in~\cite{Chatterjee:2024gje}, the allowed Hamiltonian can consist of terms with only $a_{I,j}$ or $b_{I,j^\prime}$ (but not interaction terms involving both $a_{I,j}$ and $b_{I,j^\prime}$) if locality of Hamiltonian (i.e. the ones that can be written as a sum of Hamiltonian density with local interactions) is assumed. Any interaction mixing $a_{I,j}$ and $b_{I,j^\prime}$ together is forbidden, since the distance between $a_{I,j}$ and $b_{I,j^\prime}$ involved in the interaction will keep increasing under the actions of $T$, and eventually the interaction becomes a non-local one. Furthermore, since there is fermion parity $G_2$, the allowed quadratic Hamiltonian deformations that mix together two flavors of staggered fermions, such as $a_{1,j} a_{2,j}$ or $b_{1,j} b_{2,j}$, must involve an appropriate site-dependent sign flipping phase for each local interaction. For example, $T(a_{1,j}a_{2,j})T^{-1}=-a_{1,j-1}a_{2,j-1}$, such that $T\left((-1)^j a_{1,j}a_{2,j}\right)T^{-1}=(-1)^{j-1}a_{1,j-1}a_{2,j-1}$, hence $\sum_j (-1)^j a_{1,j}a_{2,j}$ is invariant. (Similar equations hold for bilinear terms such as $(-1)^j b_{1,j^\prime} b_{2,j}$.)

Next we check whether the Hamiltonian deformations identified before are consistent with $T$ invariance. 
For this purpose, we rewrite $\Delta H_1$ and $\Delta H_2$ using Majorana fermions. According to the relations $c_{1,j}=\frac{1}{2}(a_{1,j}+i b_{1,j})$ and $c_{2,j}=\frac{1}{2}(a_{2,j}+i b_{2,j})$, we find
\begin{align}
\Delta H_1&=\sum_{j=1}^N (-1)^j c^\dagger_{1,j} c_{2,j}\nn\\
&=\sum_{j=1}^N \frac{(-1)^j}{4}\left(a_{1,j}a_{2,j}+b_{1,j}b_{2,j}+i a_{1,j} b_{2,j}-i b_{1,j} a_{2,j}\right)
\end{align}
and 
\begin{align}
\Delta H_2&=\sum_{j=1}^N\frac{(-1)^j}{2}\left(c^\dagger_{1,j}c_{2,j+1}-c^\dagger_{1,j+1} c_{2,j}\right)\nn\\
&=\sum_{j=1}^N \frac{(-1)^j}{8}\left(a_{1,j}a_{2,j+1}+ia_{1,j}b_{2,j+1}-ib_{1,j}a_{2,j+1}+b_{1,j}b_{2,j+1}\right.\nn\\
&\left.\quad\quad\quad\quad\quad\quad -a_{1,j+1}a_{2,j}+i b_{1,j+1} a_{2,j}-i a_{1,j+1}b_{2,j}-b_{1,j+1}b_{2,j}\right)\;.
\end{align}
Therefore, only the hermitian combinations 
\be
i(\Delta H_1-\text{h.c.})=\frac{i}{2}\sum_{j=1}^N (-1)^j \left(a_{1,j} a_{2,j}+b_{1,j} b_{2,j}\right)\;
\ee
and
\be
i(\Delta H_2-\text{h.c.})=\frac{i}{4}\sum_{j=1}^N (-1)^j \left(a_{1,j}a_{2,j+1}+b_{1,j}b_{2,j+1}-a_{1,j+1}a_{2,j}-b_{1,j+1}b_{2,j}\right)\;
\ee
are consistent with $T$-invariance. Instead, the other two combinations $\Delta H_1+\text{h.c.}$ and $\Delta H_2+\text{h.c.}$ are forbidden by the $T$ invariance since they contain interaction terms involving both $a_{I,j}$ and $b_{I,j}$. (For the same reason, $\Delta H_1$ and $\Delta H_2$ are not invariant under $T$ either.)

Furthermore, it is straightforward to show $i(\Delta H_1-\text{h.c.})$ and $i(\Delta H_2-\text{h.c.})$ are also invariant under infinitesimal transformation generated by $Q^V$, i.e. $\exp(i\phi Q^V)=1+i\phi Q^V+\cdots$ where $\phi\ll 1$. For the operators $O_1=a_{1,j} a_{2,j^\prime}$ and $O_2=b_{1,j} b_{2,j^\prime}$, it amounts to show that $O_1+O_2$ is invariant. 
Indeed we find that under infinitesimal transformation generated by $Q^V$,
\begin{align}
\delta O_1=\phi(b_{1,j} a_{2,j^\prime}+a_{1,j} b_{2,j^\prime})+\mathcal{O}(\phi^2)\;,\\
\delta O_2=-\phi(a_{1,j} b_{2,j^\prime}+b_{1,j} a_{2,j^\prime})+\mathcal{O}(\phi^2).
\end{align}
Hence $\delta O_1$ and $\delta O_2$ cancel each other. 

In summary, we demonstrate in this subsection that the Hamiltonian deformations in Eqs.~\eqref{eq:deformation1_model2} and~\eqref{eq:deformation2_model2} are indeed consistent with the lattice charges $Q^V$ and $Q^A$. In the continuum limit, they emanate the Dirac mass terms hence they can be used to gap the two Dirac fermions. The existence of symmetry-preserving Hamiltonian deformations on the lattice that can also gap the system implies chiral anomaly cancellation of $\mathcal{Q}^V$ and $\mathcal{Q}^A$ in the continuum, this also matches the Onsager algebra on the lattice. Notice that our results do not contradict the gaplessness constraint imposed by the lattice charges $Q^V$ and $Q^A$ in the model that flows to a single Dirac fermion~\cite{Chatterjee:2024gje}, where there is indeed a chiral anomaly between $\mathcal{Q}^V$ and $\mathcal{Q}^A$ in the continuum.

\section{A chiral fermion $3-4-5-0$ model}
\label{sec:model_3450}

Inspired by the previous results on two Dirac fermions, we define a chiral fermion $3-4-5-0$ model~\cite{Wang:2013yta, Wang:2018ugf, Zeng:2022grc, Lu:2022qtc} in $(1+1)$ dimensions using the vectorial and axial charges. It is straightforward to do so in the continuum since the chiral charges of various Weyl fermions directly follow from the linear combinations of $\mathcal{Q}^V$ and $\mathcal{Q}^A$. Given the close relations between $\mathcal{Q}^{V,A}$ and their lattice precursors $Q^{V,A}$, we attempt to define a lattice Hamiltonian model of two flavors of staggered fermions that in the continuum limit flows to the chiral fermion $3-4-5-0$ model. On top of it, we will analyze the symmetry properties (both in the continuum and on the lattice) of the multi-fermion interactions that has been proposed to drive the chiral fermions to a gapped phase without breaking the chiral symmetry~\cite{Wang:2013yta}.

In the chiral fermion $3-4-5-0$ model, no fermion bilinear mass terms are allowed by the chiral symmetry. Nevertheless, it has been proposed in~\cite{Wang:2013yta, Wang:2018ugf} that, when all 't Hooft anomalies cancel~\footnote{This includes the known and unknown 't Hooft anomalies. But in this note we focus on chiral symmetry and the perturbative chiral anomaly. In this context, SMG means a mass generation mechanism without breaking chiral symmetry (but it may break other symmetries in the theory).}, a mass gap can be generated by multi-fermion interactions that fully respect the chiral symmetry. While these interactions naively appear to be irrelevant, and thus seemingly unable to affect the infrared behavior of the theory, this intuition applies only in the regime of weak couplings. In the intermediate strongly coupled regime, these interactions become relevant and enable the generation of a mass gap for chiral fermions.
This thought-provoking proposal was later demonstrated by a numerical calculation~\cite{Zeng:2022grc}. 
In modern terminology, this mechanism where a mass gap is generated without an explicit mass term is referred to as ``symmetric mass generation'' (SMG); see e.g.~\cite{Wang:2022ucy} for a comprehensive review and also in e.g.~\cite{Razamat:2020kyf, Tong:2021phe} for some other models in various dimensions.

\subsection{In the continuum}

Similar to the previous model of two Dirac fermions, we focus on the anomaly-free and Abelian symmetry group 
\be
G^\prime_{\text{conti.}}=U(1)_V\times U(1)_A \;
\ee
of two massless and free Dirac fermions whose Hamiltonian is the same as in Eq.~\eqref{eq:ham_model2_conti}. Using these two symmetries, one can define a chiral fermion $3-4-5-0$ model in the continuum. 
The charge assignments for the various chiral fermions are listed in Table~\ref{tab:3450}, from which one can verify explicitly that all chiral anomalies between $U(1)_V$ and $U(1)_A$ cancel. 
Due to the charge assignments, the present model is a chiral one, i.e., all fermion bilinear mass terms are forbidden by the global symmetry $U(1)_V\times U(1)_A$.

\begin{table}[t]
      \centering
      \begin{tabular}{|c||c|c||c|c|c|}
      \hline
           & $U(1)_V$ & $U(1)_A$ & $U(1)_1$ & $U(1)_2$ \\
           \hline
         $\psi^\dagger_{1L}$  & $3$ & $1$ & $5$ & $4$ \\
         \hline
         $\psi^\dagger_{1R}$  & $3$  & $-1$ &  $4$ & $5$ \\
         \hline
         $\psi^\dagger_{2L}$  & $1$ & $-3$ & $0$ & $3$ \\
         \hline
         $\psi^\dagger_{2R}$  & $1$ & $3$ & $3$ & $0$ \\
         \hline
      \end{tabular}
      \caption{Charge assignments for the chiral fermion $3-4-5-0$ model in the continuum. The charge for $U(1)_1$ is given by a linear combination of vectorial and axial charges, i.e., $(3\mathcal{Q}^V+\mathcal{Q}^A)/2$.  Similarly, that for $U(1)_2$ is $(3\mathcal{Q}^V-\mathcal{Q}^A)/2$. From the charges of $U(1)_1\times U(1)_2$ it is clear that this model is the chiral fermion $3-4-5-0$ model, where explicit fermion mass terms are forbidden.  
      Notice that the same charge assignment was considered in~\cite{Lu:2022qtc}, but our lattice model differs from theirs, particularly the lattice precursor of the $U(1)_A$ symmetry is different. Here we do not rely on the Fermi surface to identify the lattice symmetry that flows to $U(1)_A$ in the continuum.}
      \label{tab:3450}
  \end{table}

It is useful to define the following charges in the continuum 
\bea
\mathcal{Q}^V &=& 3 \mathcal{Q}^{V_1}+\mathcal{Q}^{V_2}\ ,\label{eq:charge_conti_3450_1}\\
\mathcal{Q}^A &=& \mathcal{Q}^{A_1}-3 \mathcal{Q}^{A_2}\ ,
\label{eq:charge_conti_3450_2}
\eea
for $U(1)_V$ and $U(1)_A$ symmetries respectively, where $\mathcal{Q}^{V_I}$ and $\mathcal{Q}^{A_I}$ act on the corresponding Weyl fermions $(\psi_I)_{L,R}$ as in Eq~\eqref{eq:conti_action}, therefore we have
\begin{align}
[\mathcal{Q}^V, \psi^\dagger_{1,L}]&=3 \psi^\dagger_{1,L}\;, \quad 
[\mathcal{Q}^V, \psi^\dagger_{1,R}]=3\psi^\dagger_{1,R}\;, \quad
[\mathcal{Q}^V, \psi^\dagger_{2,L}]=\psi^\dagger_{2,L}\;, \quad 
[\mathcal{Q}^V, \psi^\dagger_{2,R}]=\psi^\dagger_{2,R}\; ;\nn\\
[\mathcal{Q}^A, \psi^\dagger_{1,L}]&=\psi^\dagger_{1,L}\;, \quad
[\mathcal{Q}^A, \psi^\dagger_{1,R}]=-\psi^\dagger_{1,R}\;, \quad
[\mathcal{Q}^A, \psi^\dagger_{2,L}]=-3 \psi^\dagger_{2,L}\;, \quad
[\mathcal{Q}^A, \psi^\dagger_{2,R}]=3\psi^\dagger_{2,R}\;.\nn\\
\label{eq:conti_action_model3}
\end{align}
Since all the charges commute, both fermion flavor and chirality are well-defined notions in the continuum.
By linear combinations one can define the charges for $U(1)_1\times U(1)_2$ in the continuum as
\bea
\mathcal{Q}_1 &=&\frac{1}{2}(3\mathcal{Q}^V+\mathcal{Q}^A)\;,\\
\mathcal{Q}_2 &=&\frac{1}{2}(3\mathcal{Q}^V-\mathcal{Q}^A)\;. 
\eea
Hence we have the commutators
\begin{align}
[\mathcal{Q}_1, \psi^\dagger_{1,L}]&=5 \psi^\dagger_{1,L}\;, \quad 
[\mathcal{Q}_1, \psi^\dagger_{1,R}]=4\psi^\dagger_{1,R}\;, \quad
[\mathcal{Q}_1, \psi^\dagger_{2,L}]=0 \psi^\dagger_{2,L}\;, \quad 
[\mathcal{Q}_1, \psi^\dagger_{2,R}]=3\psi^\dagger_{2,R}\; ;\nn\\
[\mathcal{Q}_2, \psi^\dagger_{1,L}]&=4\psi^\dagger_{1,L}\;, \quad
[\mathcal{Q}_2, \psi^\dagger_{1,R}]=5\psi^\dagger_{1,R}\;, \quad
[\mathcal{Q}_2, \psi^\dagger_{2,L}]=3 \psi^\dagger_{2,L}\;, \quad
[\mathcal{Q}_2, \psi^\dagger_{2,R}]=0\psi^\dagger_{2,R}\;.\nn\\
\label{eq:conti_action_model3_2}
\end{align}
These commutators are the defining equations for the chiral fermion $3-4-5-0$ model.

\subsection{On the lattice}

Again, we start from two flavors of staggered fermions whose Hamiltonian is the same as Eq.~\eqref{eq:ham_model2}. As we already see in the last section, this lattice Hamiltonian flows to a continuum QFT with two left-moving and two right-moving Weyl fermions. Motivated by Eq.~\eqref{eq:charge_conti_3450_1} and Eq.~\eqref{eq:charge_conti_3450_2}, we consider the following lattice charges:
\begin{align}
Q^V &= 3 Q^{V_1}+ Q^{V_2} \label{eq:vec_lat_charge_3450} \\
&=3 \sum_{j=1}^{N} \left(c^\dagger_{1,j} c_{1,j}-\frac{1}{2}\right)+ \sum_{j=1}^{N} \left(c^\dagger_{2,j} c_{2,j}-\frac{1}{2}\right) \\
&=\frac{i}{2} \sum_{j=1}^N (3 a_{1,j} b_{1,j}+a_{2,j} b_{2,j})\; ,
\end{align}
and
\begin{align}
Q^A&=Q^{A_1}-3 Q^{A_2}=T_{b_1} Q^{V_1} (T_{b_1})^{-1}- 3\ T_{b_2} Q^{V_2} (T_{b_2})^{-1} \label{eq:axi_lat_charge_3450} \\
&=\frac{1}{2}\sum_{j=1}^{N}(c_{1,j}+c^\dagger_{1,j})(c_{1,j+1}-c^\dagger_{1,j+1})-\frac{3}{2}\sum_{j=1}^{N}(c_{2,j}+c^\dagger_{2,j})(c_{2,j+1}-c^\dagger_{2,j+1})\\
&=\frac{i}{2}\sum_{j=1}^N (a_{1,j} b_{1,j+1}-3 a_{2,j} b_{2,j+1})\;.
\end{align}
One can evaluate the following commutator
\be
[Q^V, Q^A]= i (3 G_{1,1}-3 G_{2,1})\;. 
\label{eq:onsager_3450}
\ee
Again, the cancellation of the chiral anomaly between $U(1)_V$ and $U(1)_A$ symmetries in the continuum matches the above commutator of $[Q^V, Q^A]$, where the coefficients of $G_{1,1}$ and $G_{2,1}$ cancel each other. 

In the momentum space, the charges $Q^V$ and $Q^A$ acts on the complex lattice fermions as follows, 
\begin{align}
[Q^V, \gamma^\dagger_{1,k}] &= 3 \gamma^\dagger_{1,k}\;,\label{eq:lat_action_model_3450_1}\\
[Q^V, \gamma^\dagger_{2,k}] &=\gamma^\dagger_{2,k}\;,\label{eq:lat_action_model_3450_2}\\
[Q^A, \gamma^{\dagger}_{1,k}] &= \cos\left(\frac{2\pi }{N} k\right) \gamma^{\dagger}_{1,k} + i \sin\left(\frac{2\pi }{N} k\right) \gamma_{1,-k}\;,\label{eq:lat_action_model_3450_3}\\
[Q^A, \gamma^{\dagger}_{2,k}] &= -3\cos\left(\frac{2\pi }{N} k\right) \gamma^{\dagger}_{2,k} - 3 i \sin\left(\frac{2\pi }{N} k\right) \gamma_{2,-k}\;.
\label{eq:lat_action_model_3450}
\end{align}
With the identifications $\psi^\dagger_{I,L}(-|\epsilon|)\sim \gamma^\dagger_{I,-|\epsilon|}$ and $\psi^\dagger_{I,R}(|\epsilon|)\sim \gamma^\dagger_{I,-\frac{N}{2}+|\epsilon|}$, we observe that the actions of the lattice charges $Q^V$ and $Q^A$ coincide with the actions of $\mathcal{Q}^V$ and $\mathcal{Q}^A$ (c.f. Eq.~\eqref{eq:conti_action_model3}) in the continuum limit, where $|\epsilon|\ll N$ and $N\to\infty$. 
Likewise, we can define the lattice charges
\bea
Q_1 &=&\frac{1}{2}(3 Q^V+ Q^A)\;, \label{eq:def_charge1_3450model} \\
Q_2 &=&\frac{1}{2}(3 Q^V- Q^A)\;,  \label{eq:def_charge2_3450model}
\eea
that emanate the $\mathcal{Q}_1$ and $\mathcal{Q}_2$ in the continuum. 
Notice that $[Q_1,Q_2]=-\frac{3}{2}[Q^V, Q^A]\neq 0$ on the lattice. 
Again, the actions of microscopic lattice charges $Q_{1,2}$ agree with their continuum counterparts $\mathcal{Q}_{1,2}$ (c.f. Eq.~\eqref{eq:conti_action_model3_2}) for the low energy modes in the continuum limit. 
This is easily seen from the following commutators
\begin{align}
[Q_1, \gamma^\dagger_{1,k}] &= \frac{9}{2} \gamma^\dagger_{1,k}+ \frac{1}{2}\cos\left(\frac{2\pi }{N} k\right) \gamma^{\dagger}_{1,k} + \frac{i}{2} \sin\left(\frac{2\pi }{N} k\right) \gamma_{1,-k}\;,\\
[Q_1, \gamma^\dagger_{2,k}] &=\frac{3}{2}\gamma^\dagger_{2,k}-\frac{3}{2}\cos\left(\frac{2\pi }{N} k\right) \gamma^{\dagger}_{2,k} - \frac{3}{2} i \sin\left(\frac{2\pi }{N} k\right) \gamma_{2,-k}\;,\\
[Q_2, \gamma^{\dagger}_{1,k}] &= \frac{9}{2} \gamma^\dagger_{1,k}- \frac{1}{2}\cos\left(\frac{2\pi }{N} k\right) \gamma^{\dagger}_{1,k} - \frac{i}{2} \sin\left(\frac{2\pi }{N} k\right) \gamma_{1,-k}\;,\\
[Q_2, \gamma^{\dagger}_{2,k}] &= \frac{3}{2}\gamma^\dagger_{2,k}+\frac{3}{2}\cos\left(\frac{2\pi }{N} k\right) \gamma^{\dagger}_{2,k} + \frac{3}{2} i \sin\left(\frac{2\pi }{N} k\right) \gamma_{2,-k}
\end{align}
and the identifications between the fermions in the continuum and the low-energy modes on the lattice $\psi^\dagger_{I, L}(-|\epsilon|)\sim \gamma^\dagger_{I,-|\epsilon|}$ and $\psi^\dagger_{I, R}(|\epsilon|)\sim \gamma^\dagger_{I,-\frac{N}{2}+|\epsilon|}$.
However, since the lattice charges do not commute, we do not have the $U(1)_V\times U(1)_A$ symmetry (or the $U(1)_1\times U(1)_2$ chiral symmetry) on the lattice.  

Therefore, we observe that the lattice Hamiltonian model of two-flavor staggered fermions with the charges defined as in Eq.~\eqref{eq:vec_lat_charge_3450} and Eq.~\eqref{eq:axi_lat_charge_3450} (or the combinations in Eq.~\eqref{eq:def_charge1_3450model} and Eq.~\eqref{eq:def_charge2_3450model}) flows to the chiral fermion $3-4-5-0$ model in the continuum, where all fermions are massless.

It may be useful to compare our lattice model with earlier constructions~\cite{Wang:2013yta, Wang:2018ugf, Lu:2022qtc}. Here our construction crucially relies on the lattice charges $Q^V$ and $Q^A$. This construction is rather straightforward due to the simplicity of the correspondences between $Q^{V, A}$ on the lattice and $\mathcal{Q}^{V, A}$ in the continuum. 
On the other hand, the constructions in~\cite{Wang:2013yta, Wang:2018ugf} rely on a thought-provoking idea of decoupling the mirror sector~\cite{Eichten:1985ft} in a doubled fermionic spectrum through properly designed multi-fermion interactions. Perhaps a more similar setup is given in~\cite{Lu:2022qtc}. Like us, they do not start from a doubled fermionic spectrum. However, the difference is that they rely on the Fermi surfaces, on top of which the Fermi momenta of the low energy modes emanate the charges under the $U(1)_A$ symmetry in the continuum.

\subsection{Constructing SMG multi-fermion interactions on the lattice}

No fermion bilinear mass term is allowed in the chiral fermion $3-4-5-0$ model, as suggested by the charges under $U(1)_{1}\times U(1)_2$ in the continuum (and hence the name of the model). 
However, a mass gap for the model can be opened by the following six-fermion interaction terms~\cite{Wang:2013yta, Wang:2018ugf, Zeng:2022grc, Lu:2022qtc} that also fully respect the anomaly-free global symmetries
\begin{align}
\mathcal{O}_1 &= (\psi_{2,R}\nabla_x \psi_{2,R}) \psi_{1,R} (\psi^\dagger_{1,L} \nabla_x \psi^\dagger_{1,L}) \psi_{2,L}\; ,\\
\mathcal{O}_2 &= \psi_{2,R} (\psi^\dagger_{1,R} \nabla_x \psi^\dagger_{1,R}) \psi_{1,L} (\psi_{2,L} \nabla_x \psi_{2,L})\; .
\end{align}
Here the derivatives correspond to site splittings on the lattice, and they are necessary for identical fermions due to their anticommuting nature. 
The symmetric deformations to the free Hamiltonian in the continuum read $\delta \mathcal{H}= \int \text{d}x \ (g_1 \mathcal{O}_1+g_2 \mathcal{O}_2 + \text{h.c.})$, where $g_{1,2}$ are two undetermined coupling constants. 
Once $g_{1,2}$ exceed some critical value, $\delta \mathcal{H}$ becomes relevant~\footnote{This phase transition is understood as a Berezinskii-Kosterlitz-Thouless (BKT) transition in the bosonized Luttinger liquid model~\cite{Wang:2013yta}. However, the full phase diagram in the parameter space of $g_{1,2}$ is still unknown.} and successfully generates a mass gap for the model~\cite{Zeng:2022grc}.

It is intriguing to find the corresponding lattice operators that flow to $\mathcal{O}_{1,2}$ in the continuum limit. Here the challenge is to precisely realize the chiral structures that appeared in these interactions. For this purpose, we rewrite $\mathcal{O}_1=-(\psi_{1, R} \psi_{2, L}) (\psi^\dagger_{1, L} \psi_{2, R})^2_{\text{pt.s.}}$ and $\mathcal{O}_2=(\psi_{1, L} \psi_{2, R}) (\psi^\dagger_{1, R} \psi_{2, L})^2_{\text{pt.s.}}$, and we notice that the combination $\delta \mathcal{H}_1=\int\text{d}x (-\mathcal{O}_1+\mathcal{O}_2)$ and $\delta \mathcal{H}_2=\int \text{d}x (\mathcal{O}_1+\mathcal{O}_2)$ have the natural correspondence with the lattice operators in the form of
\be
\Delta H (\Delta j_1,\cdots, \Delta j_6) = \sum_{j=1}^N (-1)^{3j} (c_{1, j_1} c_{2, j_2})  \left( c^\dagger_{1,j_3} c_{2, j_4} \right)\left( c^\dagger_{1,j_5} c_{2, j_6} \right) \;,
\label{eq:lattice_SMG}
\ee
where the lattice sites $j_{1,\cdots,6}$ deviate from $j$ by a finite amount infinitesimal compared to $N$, i.e. $\Delta j_l= j_l-j$ and $|\Delta j_l|\ll N$ for $l=1,2,\cdots,6$ . 
Therefore $\Delta H$ is still local, though it contains non-onsite interactions. Here the site-dependent phase factor $(-1)^{3j}$ is because there are in total three right movers involved in the six-body interactions. But we will still have to specify which three are the right movers among all six fermions. 
As we justify in the following, the chirality structure in $\mathcal{O}_{1,2}$ are generated by appropriate site splittings. By performing Fourier transformation, we find in the momentum space $\Delta H (\Delta j_1,\cdots, \Delta j_6)$ reads
\begin{align}
\Delta H (\Delta j_1,\cdots, \Delta j_6)&=\sum_{k_1,\cdots, k_6} \delta_{k_1+k_2-k_3+k_4-k_5+k_6-3\frac{N}{2},0} \nn\\
&\quad\quad \text{exp}\left[\frac{2\pi i}{N}\left(k_1 \Delta j_1 + k_2 \Delta j_2 - k_3 \Delta j_3 + k_4 \Delta j_4 - k_5 \Delta j_5 + k_6 \Delta j_6 \right)\right] \nn\\
&\quad\quad \gamma_{1,k_1} \gamma_{2, k_2} \gamma^\dagger_{1, k_3} \gamma_{2, k_4}  \gamma^\dagger_{1, k_5} \gamma_{2, k_6}
\label{eq1:lattice_SMG_momentum}
\end{align}
up to an overall normalization factor which can eventually be absorbed into the coupling constants. The canonical delta enforces $k_1+k_2-k_3+k_4-k_5+k_6-3\frac{N}{2}=0$ mod $N$.

Let us first find the lattice Hamiltonian that flows to $\delta \mathcal{H}_1=\int\text{d}x (-\mathcal{O}_1+\mathcal{O}_2)$ in the continuum. 
By combining various $\Delta H (\Delta j_1,\cdots, \Delta j_6)$ with properly chosen $\{\Delta j_l\}$, the exponential factor in Eq.~\eqref{eq1:lattice_SMG_momentum} can produce a product of various projectors
\begin{align}
& \frac{1}{2} \left[ 1-\cos \left(\frac{2\pi}{N} (-k_1+k_2)\right) \right]  \cdot \frac{1}{2} \left[ 1-\cos \left(\frac{2\pi}{N} (k_3+k_4)\right) \right] \cdot \frac{1}{2} \left[ 1-\cos \left(\frac{2\pi}{N} (k_5+k_6)\right) \right] \cdot \nn\\
& \frac{1}{2} \left[ 1-\cos \left(\frac{2\pi}{N} (-k_2+k_4)\right) \right] \cdot \frac{1}{2} \left[ 1-\cos \left(\frac{2\pi}{N} (-k_2+k_6)\right) \right] \cdot e^{\frac{2\pi i}{N} [(-k_3+k_4)-(-k_5+k_6)] r}\;.
\label{eq:project_3450}
\end{align}
Indeed, e.g. for the projector
\be
\frac{1}{2} \left[ 1-\cos \left(\frac{2\pi}{N} k\right) \right]=\frac{1}{2} \left[ 1-\frac{1}{2} \text{exp}\left(\frac{2\pi i}{N} k\right) -\frac{1}{2} \text{exp}\left(-\frac{2\pi i}{N} k\right) \right]\;, 
\ee
the first term on the right-hand side implies the site splitting $\Delta j=0$, the second one implies $\Delta j=1$, and the third one implies $\Delta j=-1$, respectively. 
For the low-energy modes in the continuum limit (i.e. $k_l=0, \frac{N}{2} \ \text{mod}\ N$ for $l=1,2,\cdots,6$), we observe that $\pm k_l+k_{l^\prime}= \frac{N}{2} \ \text{mod}\ N$ when the corresponding fermions have the opposite chirality while $\pm k_l+k_{l^\prime}= 0 \ \text{mod}\ N$ when they have the same chirality. Each projector equals one when $\pm k_l+k_{l^\prime}= \frac{N}{2} \ \text{mod}\ N$ while it vanishes when $\pm k_l+k_{l^\prime}= 0 \ \text{mod}\ N$. For example, the projector $\frac{1}{2} \left[ 1-\cos \left(\frac{2\pi}{N} (-k_1+k_2)\right) \right]$ states that either $(k_1=\pm N/2, k_2=0)$ or $(k_1=0, k_2=\pm N/2)$. (The same consideration holds for the other projectors.) Therefore, in the continuum limit Eq.~\eqref{eq:project_3450} matches the fermion chirality structure 
\be
(\psi_{1,R} \psi_{2,L}) (\psi^\dagger_{1,L} \psi_{2,R}) (\psi^\dagger_{1,L} \psi_{2,R}) + (\psi_{1,L} \psi_{2,R}) (\psi^\dagger_{1,R} \psi_{2,L}) (\psi^\dagger_{1,R} \psi_{2,L}) \;,
\label{eq:ham_def_3450_op1}
\ee
which is exactly the combination needed for $\delta \mathcal{H}_1=\int\text{d}x (-\mathcal{O}_1+\mathcal{O}_2)$. Finally, the exponential factor in Eq.~\eqref{eq:project_3450} is induced by the site-splitting $r$ (where $|r|\ll N$) necessary for identical complex fermions involved in $\Delta H (\Delta j_1,\cdots, \Delta j_6)$. 
This exponential becomes an identity in the continuum limit where $(-k_3+k_4)-(-k_5+k_6)=0\ \text{mod}\ N$.

Following the same discussion above, we can find the lattice Hamiltonian that flows to $\delta \mathcal{H}_2=\int\text{d}x (\mathcal{O}_1+\mathcal{O}_2)$ in the continuum where the fermion chirality structure is given by
\be
-(\psi_{1,R} \psi_{2,L}) (\psi^\dagger_{1,L} \psi_{2,R}) (\psi^\dagger_{1,L} \psi_{2,R}) + (\psi_{1,L} \psi_{2,R}) (\psi^\dagger_{1,R} \psi_{2,L}) (\psi^\dagger_{1,R} \psi_{2,L}) \;.
\label{eq:ham_def_3450_op2}
\ee
Clearly, it corresponds to the projector induced by site splittings on the lattice as follows, 
\begin{align}
& \cos\left(\frac{2\pi i}{N} k_1\right) \cdot \nn\\
&\frac{1}{2} \left[ 1-\cos \left(\frac{2\pi}{N} (-k_1+k_2)\right) \right]  \cdot \frac{1}{2} \left[ 1-\cos \left(\frac{2\pi}{N} (k_3+k_4)\right) \right] \cdot \frac{1}{2} \left[ 1-\cos \left(\frac{2\pi}{N} (k_5+k_6)\right) \right] \cdot \nn\\
& \frac{1}{2} \left[ 1-\cos \left(\frac{2\pi}{N} (-k_2+k_4)\right) \right] \cdot \frac{1}{2} \left[ 1-\cos \left(\frac{2\pi}{N} (-k_2+k_6)\right) \right] \cdot e^{\frac{2\pi i}{N} [(-k_3+k_4)-(-k_5+k_6)] r}\;.
\label{eq:project_3450_2}
\end{align}
Here the relative minus sign between two terms in Eq.~\eqref{eq:ham_def_3450_op2} matches $\cos\left(\frac{2\pi i}{N} k_1\right) $, while the rest projectors work the same as in the previous case. 

\subsection{Symmetry properties of the lattice SMG multi-fermion interactions}

Finally, we would like to understand whether the Hamiltonian deformation in Eq.~\eqref{eq:lattice_SMG} (or equivalently Eq.~\eqref{eq1:lattice_SMG_momentum} in momentum space) is fully compatible with the lattice charges $Q^V$ and $Q^A$. 

By evaluating the following commutator 
\be
[Q^V, \Delta H (\Delta j_1,\cdots, \Delta j_6)]= (-3-1+3-1+3-1) \cdot \Delta H (\Delta j_1,\cdots, \Delta j_6) =0\;, 
\ee
we find $\Delta H (\Delta j_1,\cdots, \Delta j_6)$ is fully compatible with the lattice charge $Q^V$ for any choice of $\{\Delta j_l\}$ with $|\Delta j_l|\ll N$. On the other hand, it is less trivial to understand whether $\Delta H (\Delta j_1,\cdots, \Delta j_6)$ can lead to a combination fully compatible with $Q^A$ as well, even though it is indeed invariant under $\mathcal{Q}^A$ in the continuum. This can be easily checked with the commutators in the momentum space, 
\begin{align}
&[Q^A, \gamma_{1,k_1} \gamma_{2,k_2} \gamma^\dagger_{1,k_3} \gamma_{2,k_4} \gamma^\dagger_{1,k_5} \gamma_{2,k_6}] \nn\\
&=\left[ -\cos\left(\frac{2\pi}{N} k_1\right) + 3 \cos\left(\frac{2\pi}{N} k_2\right) + \cos\left(\frac{2\pi}{N} k_3\right) +3 \cos\left(\frac{2\pi}{N} k_4\right) +\cos\left(\frac{2\pi}{N} k_5\right) +3\cos\left(\frac{2\pi}{N} k_6\right) \right] \nn\\
&\cdot \gamma_{1,k_1} \gamma_{2,k_2} \gamma^\dagger_{1,k_3} \gamma_{2,k_4} \gamma^\dagger_{1,k_5} \gamma_{2,k_6} +\cdots\;, 
\label{eq:QA_3450_SMG_check}
\end{align}
where we have dropped the contribution proportional to $\sin(2\pi k_l/N)$ which vanishes when $k_l=0$ or $k_l=\pm \frac{N}{2}$. For the various $\cos(2\pi k_l/N)$ terms, we find that Eq.~\eqref{eq:QA_3450_SMG_check} does not vanish exactly but it vanishes when 
\be
k_1=\pm \frac{N}{2}, \; k_2=0, \; k_3=0, \; k_4=\pm \frac{N}{2}, \; k_5=0,\; k_6=\pm\frac{N}{2}, 
\ee
and
\be
k_1=0, \; k_2=\pm\frac{N}{2}, \; k_3=\pm\frac{N}{2}, \; k_4=0, \; k_5=\pm\frac{N}{2},\; k_6=0 .
\ee
Indeed they match the fermion chirality patterns in $\mathcal{O}_{1,2}$ in the continuum.
Here $\mathcal{Q}^A$ emanates from $Q^A$ and generates an exact symmetry only in the continuum QFT at low energies. 
The cancellation of chiral anomaly signaled by the Onsager algebra in Eq.~\eqref{eq:onsager_3450} implies that there should not be obstructions for the SMG interactions in the continuum. (A priori, this does not mean that there should not be obstructions from the lattice charges, however.)

To have a better understanding, let us construct the translation operator 
\begin{align}
T&=e^{-i\frac{\pi}{2} Q^A} e^{i\frac{\pi}{2} Q^V}
=e^{-i\frac{\pi}{2} Q^{A_1}} e^{i\frac{3\pi}{2} Q^{A_2}} e^{i\frac{3\pi}{2} Q^{V_1}} e^{i\frac{\pi}{2} Q^{V_2}}\nn\\
&=\left(e^{-i\frac{\pi}{2} Q^{A_1}} e^{i\frac{\pi}{2} Q^{V_1}}\right) e^{i \pi Q^{V_1}} \left(e^{-i\frac{\pi}{2} Q^{A_2}} e^{i\frac{\pi}{2} Q^{V_2}} \right)\nn\\
&= (T_{b_1} T^{-1}_{a_1}) G_1 (T_{b_2} T^{-1}_{a_2})
\label{eq:latticeTrans_model3450}
\end{align}
where we have used the fact that $Q^{A_2}$ is a quantized charge~\cite{Chatterjee:2024gje} such that $e^{i\frac{3\pi}{2} Q^{A_2}}\sim e^{-i\frac{\pi}{2} Q^{A_2}}$, and $G_1$ is the fermion parity operator acting on the lattice Majorana fermions $a_{1,j}$ and $b_{1, j}$ as $G_1 a_{1,j} (G_1)^{-1}=-a_{1,j}$ and $G_1 b_{1,j} (G_1)^{-1}=-b_{1,j}$ while it leaves $a_{2,j}$ and $b_{2,j}$ invariant. Inspired by Eq.~\eqref{eq:lattice_SMG}, we consider the $T$-invariant six-fermion Hamiltonian 
\be
H_{1,T} (\Delta j_1,\cdots, \Delta j_6)=\sum_j (-1)^{3j} a_{1,j_1} a_{1,j_3} a_{1,j_5} a_{2, j_2} a_{2, j_4} a_{2, j_6}
\ee
or
\be
H_{2,T} (\Delta j_1,\cdots, \Delta j_6) =\sum_j (-1)^{3j} b_{1,j_1} b_{1,j_3} b_{1,j_5} b_{2, j_2} b_{2, j_4} b_{2, j_6}\;,
\ee
where the site-dependent phase factor $(-1)^{3j}$ matches the action of the fermion parity operator $G_1$ on the lattice Majorana fermions (i.e. either three $a_{1,j}$ or three $b_{1,j}$) involved in the six-fermion interactions. Notice that all the interactions mixing together $a_{I,j}$'s and $b_{I^\prime, j^\prime}$'s must be forbidden by $T$-invariance if the locality of the Hamiltonian deformation is assumed. 

Under infinitesimal transformation generated by the lattice charge $Q^V$ ( i.e. $\exp(i\phi Q^V)=1+i\phi Q^V+\cdots$ where $\phi\ll 1$), neither $H_{1,T}$ nor $H_{2,T}$ is invariant and they cannot cancel each other. However, since $\Delta H (\Delta j_1,\cdots, \Delta j_6)$ is fully compatible with the lattice charge $Q^V$ for any choice of $\{\Delta j_l\}$ with $|\Delta j_l|\ll N$, one cannot obtain either $H_{1,T}$ or $H_{2,T}$ from $\Delta H (\Delta j_1,\cdots, \Delta j_6)$ with all the other terms mixing $a_{I,j}$'s and $b_{I^\prime, j^\prime}$'s being canceled out. 
We note that $H_{1,T}$ and $H_{2,T}$ may reduce to simpler forms of four-fermion or two-fermion interactions when some lattice sites of Majorana fermions are identical.~\footnote{For example, $H_{1,T}$ reduces to $\sum_j (-1)^{3j} a_{1,j_1} a_{1,j_3} a_{1,j_5} a_{2, j_2}$ when $j_4=j_6$ and further to $\sum_j (-1)^{j} a_{1,j_5} a_{2, j_2}$ when $j_1=j_3$. Similar reductions can happen to $H_{2,T}$ as well.} However, there is still no combination that can be invariant under infinitesimal transformation generated by $Q^V$. In particular, due to the chiral nature of the model, Majorana fermion bilinears are not invariant, either.

In summary, we conclude that the SMG six-fermion interactions are compatible with $\mathcal{Q}^V$ and $\mathcal{Q}^A$ in the continuum, while they are only compatible with $Q^V$ on the lattice. We note that even though the SMG six-fermion interactions violate the lattice charge $Q^A$~\footnote{Perhaps an intriguing possibility is that one can further refine the lattice charges such that they preserve the algebraic structure characterizing the chiral anomalies while allowing for more symmetric deformations.}, the $U(1)_A$ symmetry generated by $\mathcal{Q}^A$ in the continuum is still an exact symmetry, hence this is still consistent with the paradigm of SMG in continuum QFT.

\section{Conclusion and outlook}
\label{sec:conclusion}

There has been some interest in recent literature to identify the microscopic origins in UV lattice models for various global symmetries and anomalies in continuum QFT in the IR. 
The surprise is that there are still new results on even the ordinary chiral symmetries and chiral anomalies in $(1+1)$ dimensions for staggered fermions~\cite{Chatterjee:2024gje}. 

In this note, we take one step further and analyze the quantized charges in lattice Hamiltonian models of multi-flavor staggered fermions that in the continuum limit emanate an anomaly-free and Abelian global symmetry group for multi Dirac fermions. (Note that the IR symmetries emanate from UV symmetries, and they are exact symmetries in the continuum QFT.) Two concrete examples are analyzed in detail in Section~\ref{sec:model_2} and~\ref{sec:model_3450}. 
In a lattice model that flows to two Dirac fermions, we identify quadratic Hamiltonian deformations that can gap the system while fully preserving both the vectorial and axial charges on the lattice. These deformations flow to the usual symmetry-preserving Dirac mass terms in the continuum. Additionally, we propose a lattice model that flows to the chiral fermion $3-4-5-0$ model in the continuum by using these lattice charges, and we construct the SMG six-fermion interactions on the lattice in our setup and analyze their symmetry properties. 

Here are some future directions.
\begin{enumerate}

\item In scenarios without chiral anomalies in the continuum, there is no obstruction to coupling both $\mathcal{Q}^V$ and $\mathcal{Q}^A$ to gauge fields, it may be interesting to understand the implications on the lattice when both $Q^V$ and $Q^A$ are coupled to gauge fields~\cite{Berkowitz:2023pnz}, and perhaps to consider the interplays with generalized symmetries and anomalies such as~\cite{Honda:2022edn, Berkowitz:2023pnz}. It might also be interesting to consider interplays with other simple symmetry charges and anomalies such as CRT anomaly on the lattice in various dimensions~\cite{Li:2024dpq}.  

\item For theories without chiral anomalies, can we construct on the lattice a boundary such that it flows to a symmetry-preserving boundary in the continuum? This conceptually is similar to finding concrete lattice constructions for the lattice Hamiltonian deformations that in the continuum limit flow to symmetry-preserving interactions. Can the lattice charges $Q^V$ and $Q^A$ play a role here? If this is achieved in the chiral fermion $3-4-5-0$ model, one might be able to numerically simulate a toy model closely related to the fermion-monopole scattering problem~\cite{vanBeest:2023dbu, vanBeest:2023mbs}.

\item Once the Onsager algebra signals chiral anomaly cancellation, it follows that there is no obstruction to having symmetry-preserving deformations in the continuum that generate the mass gap. However, this does not automatically imply that the correponding lattice Hamiltonian deformations are fully compatible with the lattice charges. 

We observe that, in the chiral fermion $3-4-5-0$ model, the SMG six-fermion interactions are not compatible with the lattice charge $Q^A$ (but they are compatible in the continuum limit). Perhaps this calls for a more general definition for the lattice charge $Q^A$, whose algebra ideally still captures the chiral anomaly while allowing for symmetric deformations fully aligned with the continuum.~\footnote{Nonetheless, perhaps we should not feel too surprised. After all, as we already mentioned in the introduction, the Hamiltonian deformations that are consistent with $Q^{V, A}$ on the lattice must also be invariant under the $U(1)_V\times U(1)_A$ symmetry in the continuum, but not necessarily the other way around.}

\item Finally, one may pursue further generalizations to analyze lattice charges that emanate non-Abelian charges in the continuum and in higher dimensions. From this perspective, we hope to revisit 't Hooft anomalies and the construction of the equations of persistent mass condition for the usual chiral symmetries in QCD on the lattice; see recently~\cite{Ciambriello:2022wmh, Ciambriello:2024xzd, Ciambriello:2024msu} for some new insights on this topic in the continuum. 

\end{enumerate}

\begin{acknowledgments}
We would like to thank Yahui Chai for many helpful discussions on staggered fermions, and Andrea Luzio for valuable insights and collaboration at the early stage. 
The work is partially supported by European Research Council (ERC) grant n.101039756.
We are grateful to the Mainz Institute for Theoretical Physics (MITP) of the Cluster of Excellence PRISMA$^+$ (Project ID 390831469), for its hospitality and support during the preparation of this work. 
\end{acknowledgments}


\begin{appendix}

\section{Naive discretization and fermion doubling}
\label{app1}

Given the Lagrangian or Hamiltonian in the continuum, one may attempt to find the corresponding lattice model by naive discretization.  

Let us focus on the left-moving Weyl fermion and naively discretize the field, the Hamiltonians in the continuum and on the lattice are respectively 
\bea
\mathcal{H}_L &=& i \int \text{d}x \ \psi^\dagger_L \partial_x \psi_L  \;,\\ 
H_L &=& i \sum^{N}_{j=1} \frac{1}{2} \psi^\dagger_{L,j} \left(\psi_{L,j+1}-\psi_{L,j-1}\right) =\frac{i}{2} \sum^{N}_{j=1} (\psi^\dagger_{L,j} \psi_{L,j+1} + \psi_{L,j} \psi^\dagger_{L,j+1} )\ ,
\label{eq:naivedisc}
\eea
where the spatial lattice is a closed chain with in total $N$ sites.  
The complex lattice fermions $\psi_{L,j}$ satisfy the Clifford algebra $\{\psi_{L,j}, \psi^\dagger_{L,j^\prime}\}=\delta_{j,j^\prime}, \{\psi_{L,j}, \psi_{L,j^\prime}\}=\{\psi^\dagger_{L,j}, \psi^\dagger_{L,j^\prime}\}=0$.
By using the Fourier transformation on the lattice, we rewrite $\psi_{L,j}$ as
\be
\psi_{L,j}=\frac{1}{\sqrt{N}}\sum_{k} e^{\frac{2\pi i}{N} k j } \psi_{L,k}, 
\ee
where $\psi_{L,k}$ satisfy 
$\{\psi_{L,k},\psi^\dagger_{L,k^\prime}\}=\delta_{k,k^\prime}$ and $\{\psi_{L,k},\psi_{L,k^\prime}\}=\{\psi^\dagger_{L,k},\psi^\dagger_{L,k^\prime}\}=0$.

In the momentum space the Hamiltonian in Eq.~\eqref{eq:naivedisc} corresponds to
\be
H_L=\frac{i}{2N} \sum_{k,k',j} e^{\frac{2\pi i}{N} (-k+k') j}\left(e^{\frac{2\pi i}{N} k'}-e^{-\frac{2\pi i}{N} k'} \right) \psi^{\dagger}_{L,k} \psi_{L,k'} = -\sum_{k} \sin\left(\frac{2\pi}{N} k\right) \psi^\dagger_{L,k} \psi_{L,k}\;.
\ee
Since $k\sim k+N$ it is sufficient to consider $k\in [-\frac{N}{2},\frac{N}{2})$. Taking the continuum limit amounts to focusing on the low-energy modes near $k=0$ and $k=\pm \frac{N}{2}$ and then taking the limit $N\to \infty$.
We see explicitly that, by naive discretization in terms of $H_L$, there is one left-moving Weyl fermion near $k=0$ and one right-moving Weyl fermion near $k=\pm\frac{N}{2}$. This is the infamous fermion doubling problem on the lattice, i.e. a low-energy left-mover is intrinsically related to another right-mover. Because of this, defining fermion chirality on the lattice is a challenging task.

Likewise, if we discretize $\mathcal{H}_R = -i \int \text{d}x \ \psi^\dagger_R \partial_x \psi_R$, we obtain 
\be
H_R=-\frac{i}{2} \sum^{N}_{j=1} (\psi^\dagger_{R,j} \psi_{R,j+1} + \psi_{R,j} \psi^\dagger_{R,j+1} )=\sum_{k} \sin\left(\frac{2\pi}{N} k\right) \psi^\dagger_{R,k} \psi_{R,k}\;, 
\ee
hence there is one right-moving Weyl fermion near $k=0$ and another left-moving Weyl fermion near $k=\pm\frac{N}{2}$. By the field redefinition on the lattice $\psi^\prime_{R,j}=(-1)^j \psi_{R,j}$ (and equivalently $\psi^\prime_{R,k}=\psi_{R,k+\frac{N}{2}}$ in the momentum space), $H_R$ can be rewritten as
\be
H_R=\frac{i}{2} \sum^{N}_{j=1} (\psi^{\prime\dagger}_{R,j} \psi^\prime_{R,j+1} + \psi^\prime_{R,j} \psi^{\prime\dagger}_{R,j+1} )=-\sum_{k} \sin\left(\frac{2\pi}{N} k\right) \psi^{\prime\dagger}_{R,k} \psi^\prime_{R,k}\;,
\ee
which has exactly the same form as $H_L$. Again, this suggests that fermion chirality is not well-defined on the lattice. 

Compared to the naive discretization, the staggered fermion approach~\cite{Kogut:1974ag, Banks:1975gq, Susskind:1976jm} effectively reduces the total degrees of freedom by one-half. The complex lattice fermion $c_j$ in Eq.~\eqref{eq:Ham_staggered} are not viewed as the naive discretization of either $\psi_L$ or $\psi_R$ in the continuum, although the Hamiltonian looks the same. Instead, $c_j$ on odd and even sites on the lattice correspond to upper and lower components of a Dirac fermion in the continuum~\cite{Susskind:1976jm} (which is defined in a different basis from left and right movers). Hence both upper and lower components are mixtures of the left and right moving fermions $\psi_L$ and $\psi_R$.

\section{Actions of the quantized charges and the Onsager algebra}
\label{app2}

Throughout the paper, we have been relying on the actions of the lattice charges on the Majorana and complex lattice fermions. In this appendix, we review the commutation relations that are useful for deriving the results in the paper.

With the lattice charges 
\be
Q^{V_I}=\sum_{j=1}^N \left(c^\dagger_{I,j} c_{I,j}-\frac{1}{2}\right)=\frac{i}{2} \sum_{j=1}^N a_{I,j} b_{I,j}
\ee
and 
\be
Q^{A_I}=\frac{1}{2}\sum_{j=1}^N \left(c_{I,j} +c^\dagger_{I,j}\right)\left(c_{I,j+1}-c^\dagger_{I,j+1}\right)=\frac{i}{2} \sum_{j=1}^N a_{I,j} b_{I,j+1}\;,
\ee
it is straightforward to determine their actions on the lattice fermions. For the lattice Majorana fermions, we find 
\be
[Q^{V_I}, a_{I^\prime, j}]=-i b_{I, j} \delta_{I,I^\prime}\;,\quad\quad [Q^{V_I}, b_{I^\prime, j}]= i a_{I, j} \delta_{I,I^\prime}\;.
\label{eq:app2_vec}
\ee
and 
\be
[Q^{A_I}, a_{I^\prime, j}]=-i b_{I, j+1} \delta_{I,I^\prime}\;,\quad\quad [Q^{A_I}, b_{I^\prime, j}]= i a_{I, j-1} \delta_{I,I^\prime}\;.
\label{eq:app2_axi}
\ee

Likewise, one can work out the commutators for the complex lattice fermions $c_{I,j}$ using the relation $c_{I,j}=\frac{1}{2} (a_{I,j}+i b_{I,j})$. After Fourier transformation, one obtains the corresponding commutators for the complex lattice fermions in the momentum space, e.g., the equations from Eq.~\eqref{eq:lat_action_model2_1} to Eq.~\eqref{eq:lat_action_model2_4} in the vectorlike model of two Dirac fermions, and the equations from Eq.~\eqref{eq:lat_action_model_3450_1} to Eq.~\eqref{eq:lat_action_model_3450} in the chiral fermion $3-4-5-0$ model. These commutators in the momentum space are particularly useful for taking the low-energy continuum limit. 

From Eqs.~\eqref{eq:app2_vec} and~\eqref{eq:app2_axi}, one can deduce the finite actions of the lattice charges on the lattice Majorana fermions. In particular, we find
\begin{align}
e^{i\phi Q^{V_I}} a_{I^\prime, j} e^{-i\phi Q^{V_I}} &= \cos(\phi \delta_{I,I^\prime}) \ a_{I^\prime, j} + \sin(\phi \delta_{I,I^\prime}) \ b_{I^\prime, j}\;, \\
e^{i\phi Q^{V_I}} b_{I^\prime, j} e^{-i\phi Q^{V_I}} &= \cos(\phi \delta_{I,I^\prime}) \ b_{I^\prime, j} - \sin(\phi \delta_{I,I^\prime}) \ a_{I^\prime, j}\;, \\
e^{i\phi Q^{A_I}} a_{I^\prime, j} e^{-i\phi Q^{A_I}} &= \cos(\phi \delta_{I,I^\prime}) \ a_{I^\prime, j} + \sin(\phi \delta_{I,I^\prime}) \ b_{I^\prime, j+1}\;, \\
e^{i\phi Q^{A_I}} b_{I^\prime, j} e^{-i\phi Q^{A_I}} &= \cos(\phi \delta_{I,I^\prime}) \ b_{I^\prime, j} - \sin(\phi \delta_{I,I^\prime}) \ a_{I^\prime, j-1}\;,
\end{align}
where $\phi\sim \phi+2\pi$.
These finite actions are particularly useful in constructing the lattice translation operators
in Eqs.~\eqref{eq:latticeTrans_model2},~\eqref{eq:latticeTrans_model3450} and understanding their actions on $a_{I, j}, b_{I, j}$.

Following~\cite{Chatterjee:2024gje}, we define $Q_{I,n}=\frac{i}{2} \sum_{j=1}^N a_{I,j} b_{I, j+n}$, where $n=0$ and $n=1$ correspond to the lattice charges in Eqs.~\eqref{eq:app2_vec} and~\eqref{eq:app2_axi}. It is straightforward to obtain the following Onsager algebra for the multi-fermion case
\begin{align}
[Q_{I,n}, Q_{I^\prime, n^\prime}] &= i \ G_{I,n^\prime-n} \ \delta_{I, I^\prime} \;, \label{eq:app2_Onsagar1}\\
[Q_{I,n}, G_{I^\prime, n^\prime}] &= 2 i (Q_{I, n-n^\prime}-Q_{I, n+n^\prime})\ \delta_{I, I^\prime} \;,\\
[G_{I,n},G_{I^\prime, n^\prime}] &= 0\;,
\end{align}
where $G_{I,n}=\frac{i}{2}\sum_j(a_{I,j}a_{I,j+n}-b_{I,j}b_{I,j+n})$ whose matrix elements between any low-energy states vanish in the continuum limit.
From Eq.~\eqref{eq:app2_Onsagar1}, one can obtain Eq.~\eqref{eq:Latticeanomaly_2Dirac} in the vectorlike model of the two Dirac fermions and Eq.~\eqref{eq:onsager_3450} in the chiral fermion $3-4-5-0$ model, which match the perturbative chiral anomaly in the continuum QFT. 

\end{appendix}

\bibliography{notes.bib}

\end{document}